\documentclass[12pt]{article}
\usepackage{amssymb} 
\usepackage{hhline}
\usepackage[scale={.75,.75}]{geometry}
\usepackage[T1]{fontenc}
\usepackage{bbm} 
\usepackage{fleqn} 
\usepackage{mathrsfs} 
\usepackage{cite} 
\usepackage{longtable}
\usepackage{caption}
\usepackage{axodraw}
\usepackage{cite}

\usepackage{axodraw}
\usepackage{color}
\usepackage{amsmath}
\usepackage{mathrsfs}
\usepackage{verbatim}
\usepackage{graphicx}
\usepackage{upgreek}
\DeclareMathAlphabet{\mathpzc}{OT1}{pzc}{m}{it}	

\usepackage{lscape}
\usepackage{psfrag}

\newcommand{\e}{\epsilon}

\newcommand{\q}{\bf q}
\newcommand{\Boxq}{(\partial_{t_1}^2+\omega_{{\bf q}}^2)}

\newcommand{\be}{\begin{equation}}
\newcommand{\ee}{\end{equation}}
\newcommand{\bea}{\begin{array}}
\newcommand{\eea}{\end{array}}
\newcommand{\Dm}{\Delta^-_{\q}}

\newcommand{\bes}{\begin{split}}
\newcommand{\ees}{\end{split}}
\newcommand{\Ccnt}{{\mathcal C}}
\newcommand{\tp}{t}
\newcommand{\tm}{y}
\newcommand{\Lm}{{\mathcal L}}
\newcommand{\Binom}[2]{B^{#1}_{#2}}

\DeclareMathOperator{\re}{Re}
\DeclareMathOperator{\im}{Im}

\DeclareMathOperator{\Tr}{Tr}

\numberwithin{equation}{section}
\numberwithin{table}{section}
\allowdisplaybreaks

\begin{document}

\title{ 
{\normalsize     
DESY 08-124\hfill\mbox{}\\
December 2008\hfill\mbox{}}\\
\vspace{2cm}
\textbf{Nonequilibrium Dynamics of \\Scalar Fields in a Thermal Bath} \\[8mm]}
%
\author{A.~Anisimov, W.~Buchm\"uller, M.~Drewes, S.~Mendizabal\\[2mm]
{\normalsize\it Deutsches Elektronen-Synchrotron DESY, Hamburg, Germany}
}
\date{\mbox{}}
\maketitle

\thispagestyle{empty}


\begin{abstract}
\noindent
We study the approach to equilibrium for a scalar field which is coupled to 
a large thermal bath. Our analysis of the initial value problem is based 
on Kadanoff-Baym equations which are shown to be equivalent to a stochastic
Langevin equation. The interaction with the thermal bath generates a
temperature-dependent spectral density, either through decay and inverse decay 
processes or via Landau damping. In equilibrium, energy density and pressure
are determined by the Bose-Einstein distribution function evaluated at a
complex quasi-particle pole. The time evolution of the statistical propagator 
is compared with solutions of the Boltzmann equations for particles as well as 
quasi-particles. The dependence on initial conditions and the range of validity
of the Boltzmann approximation are determined.
\end{abstract}

\newpage\index{}

\section{Introduction}

The current standard model of cosmology explains many features of our 
universe as the result of out-of-equilibrium processes during its very early
high-temperature phase (cf.~\cite{kt90,muk05}). This includes the 
matter-antimatter asymmetry, i.e. the origin of matter, the production of dark 
matter, the formation of light elements and the decoupling of photons leading 
to the cosmic microwave background.

Many nonequilibrium processes in the early universe can be treated in the
canonical way by means of Boltzmann equations (cf.~\cite{kt90}) with 
sufficient accuracy. In some cases, however, quantum effects play a crucial 
role. This applies in particular to baryogenesis, the generation of the 
matter-antimatter asymmetry. Here the CP asymmetry, which leads to the 
baryon asymmetry, is the result of a quantum interference. It is therefore
important to go beyond the classical Boltzmann equations and to treat
the entire baryogenesis process quantum mechanically.

An attractive baryogenesis scenario is leptogenesis \cite{fy86,lut92}, where 
a quantitative understanding of the baryon asymmetry in terms of neutrino 
properties has been achieved \cite{bpy05}. In leptogenesis the 
out-of-equilibrium dynamics of a heavy Majorana neutrino, which is coupled
to a large thermal bath of standard model particles, is the origin of the
baryon asymmetry. Given the simplicity of this process, a full quantum 
mechanical treatment may be possible and some progress in this direction has 
already been made during the past years \cite{bf00,lm05,dsr07}. One
important application is the study of flavor effects \cite{bcx99}.

The treatment of nonequilibrium processes in quantum field theory is usually
based either on Kadanoff-Baym equations and the Schwinger-Keldysh formalism
\cite{kb62,kel64,leb96,ber04} or on stochastic Langevin equations 
\cite{csx84,zin93,yok04,bdh04}. Both methods have been applied to various processes 
in particle physics and cosmology, including also electroweak baryogenesis
\cite{kpx01}. In this paper we examine the connection 
between both approaches, which has been also considered in \cite{greiner}. As we shall see, the Kadanoff-Baym equations and the Langevin equation are, in fact, equivalent for
the case of a large thermal bath where backreaction effects can be neglected. 

Boltzmann equations are first-order differential equations for number 
densities, which are local in time. They represent a valuable approximation
for nonequilibrium processes in a dilute, weakly coupled gas. However, when 
the interactions between the quanta of the thermal plasma are strong, which is 
certainly the case in the presence of non-Abelian gauge interactions, the 
validity of the Boltzmann approximation is questionable. Correspondingly,
the notion of number density becomes ambiguous, although several useful 
definitions have been suggested \cite{ber04,bdh04}.  

In this paper we study the approach to equilibrium for a scalar field 
which is coupled to a thermal bath with many degrees of freedom such that
backreaction effects can be neglected. We shall focus on the description of 
this nonequilibrium process in terms of Green's functions rather than
number densities. This is analogous to studies of preheating after inflation 
based on the statistical propagator \cite{ber04,brs08}. As we shall see,
the Kadanoff-Baym equations and the Langevin equation lead to identical
results.

Knowing the exact solution of the initial value problem for the Green's
function of the scalar field, we can systematically study the conditions
for the validity of ordinary Boltzmann equations as well as Boltzmann
equations for quasi-particles. At large times the scalar field reaches
equilibrium. As we shall see, this state does not correspond to a gas
of quasi-particles. There is an additional thermal `vacuum' contribution
which in principle can even lead to a negative pressure of low-momentum
modes. The general solution of the Green's function also allows us to
study the dependence of the equilibration on the initial conditions. This
is an important problem in leptogenesis, because the
baryon asymmetry can only be predicted in terms of neutrino properties when there is no dependence
on the initial conditions \cite{bdp02}.

To illustrate our results we consider a toy model of three scalars \cite{wel83,bdh04,dre06}, one being much heavier than the 
other two.
Two particles are in thermal equilibrium whereas the third one slowly 
approaches thermal equilibrium starting from zero initial abundance. Due
to the interaction with the thermal bath this particle has a non-trivial
spectral density, approximately described by a `thermal mass' and a
`thermal width'. These are generated either by decays and inverse decays
or by a process similar to Landau damping. Some aspects of this model have  
previously been studied based on the time evolution of a number density
\cite{bdh04}.

The paper is organized as follows. In Section~2 we define the various
 Green's functions in the Schwinger-Keldysh formalism and present a brief derivation 
of the Kadanoff-Baym equations. The theoretical framework leading to the 
Langevin equation is discussed in Section~3, following \cite{bdh04}.
Section~4 deals with the solutions of the Kadanoff-Baym equations.
Thermal equilibrium and the quasi-particle picture are discussed in Section~5,
and a sytematic comparison with Boltzmann equations is made in Section~6.
The results are illustrated for a thermal bath of scalars in Section~7.
A brief summary and outlook is given in Section~8. Various properties
of the spectral function are discussed in the Appendix.

\section{The Schwinger-Keldysh formalism}

Let us consider the nonequilibrium dynamics of a scalar field. In the
Schwinger-Keldysh formalism the basic quantity is the Green's function defined on a
contour $C$ in the complex $x^0$-plane (cf.~Figure~1), 
\begin{equation}\label{causal}
\Delta_C(x_1,x_2)
=\theta_C(x^0_1,x^0_2)\Delta^>(x_1,x_2) 
+ \theta_C(x^0_2,x^0_1)\Delta^<(x_1,x_2)\ .
\end{equation}
The $\theta$-functions enforce path ordering along the contour $C$, and 
$\Delta^>$ and $\Delta^<$ are the correlation functions
\begin{align}
\Delta^>(x_1,x_2) = \langle \Phi(x_1)\Phi(x_2)\rangle =
\Tr(\uprho\Phi(x_1)\Phi(x_2))\;,\label{forw}\\ 
\Delta^<(x_1,x_2) = \langle \Phi(x_2)\Phi(x_1)\rangle =
\Tr(\uprho\Phi(x_2)\Phi(x_1))\label{back}\ , 
\end{align}
where $\uprho$ is the density matrix of the system at some initial time $t_i$.

We consider the case that the field $\Phi$ is coupled to a thermal bath 
described by a self-energy $\Pi$. The Green's function $\Delta_C$ 
then satisfies the Schwinger-Dyson equation
\begin{equation}\label{sde}
(\square_1 +m^2)\Delta_C(x_{1},x_{2})+\int_{C}d^{4}x' \Pi_{C}(x_{1},x')
\Delta_{C}(x',x_{2})=-i\delta_{C}(x_{1}-x_{2})\ ,
\end{equation}
where $\Box_1=(\partial^2/\partial x_1^2)$. Like the Green's function, also 
the self-energy can be decomposed as
\begin{equation}
\Pi_C(x_1,x_2)
=\theta_C(x^0_1,x^0_2)\Pi^>(x_1,x_2) + \theta_C(x^0_2,x^0_1)\Pi^<(x_1,x_2)\ .
\end{equation}

In the Schwinger-Dyson equation the time coordinates of $\Delta_C$ and $\Pi_C$
can be on the upper or lower branch of the contour $C$, which we denote by
the subscripts `$+$' and `$-$', respectively. Obviously, one has
\begin{align}
\Delta_{-+}(x_1,x_2) &= \Delta^>(x_1,x_2)\ , \quad 
\Delta_{+-}(x_1,x_2) = \Delta^<(x_1,x_2)\ ,\\
\Pi_{-+}(x_1,x_2) &= \Pi^>(x_1,x_2)\ , \quad 
\Pi_{+-}(x_1,x_2) = \Pi^<(x_1,x_2)\ ,
\end{align}
whereas $\Delta_{++}$, $\Pi_{++}$ and  $\Delta_{--}$, $\Pi_{--}$ are
causal and anti-causal Green functions, respectively.
From the Schwinger-Dyson equation (\ref{sde}) one obtains for the 
correlation functions $\Delta^<$ and $\Delta^>$,
\begin{align}
(\square_1+m^2)\Delta^<(x_1,x_2) &= 
\int d^{4}x'\left(-\Pi_{++}(x_1,x')\Delta^<(x',x_2) + 
\Pi^<(x_1,x')\Delta_{--}(x',x_2)\right)\ , \label{back1}\\
(\square_1+m^2)\Delta^>(x_1,x_2) &=
\int d^{4}x'\left(-\Pi^>(x_1,x')\Delta_{++}(x',x_2) + 
\Pi_{--}(x_1,x')\Delta^>(x',x_2)\right)\ , \label{forw1}
\end{align}
where the relative sign in the integrands is due to the anti-causal
time ordering on the lower branch of $C$.

\begin{figure}[t]
  \centering
    \includegraphics[width=12cm]{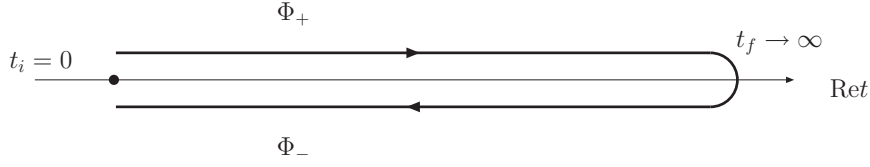}
    \caption{Path in the complex time plane for nonequilibrium Green's 
functions.\label{contour1}}
\end{figure}

It is convenient to also introduce retarded and advanced Green functions,
\begin{align}
\Delta^{R}(x_{1},x_{2})
&=\theta(t_{1}-t_{2})(\Delta^{>}(x_{1},x_{2})
-\Delta^{<}(x_{1},x_{2}))\label{retarded}\\
&=\theta(t_{1}-t_{2})\langle[\phi(x_{1}),\phi(x_{2})]\rangle \nonumber\\
&=\Delta_{++}(x_1,x_2)-\Delta_{+-}(x_1,x_2) \nonumber\\ 
&=\Delta_{-+}(x_1,x_2)-\Delta_{--}(x_1,x_2)\ ,\nonumber\\
\Delta^{A}(x_{1},x_{2})
&=-\theta(t_{2}-t_{1})(\Delta^{>}(x_{1},x_{2})
-\Delta^{<}(x_{1},x_{2}))\label{advanced}\\
&=-\theta(t_{2}-t_{1})\langle[\phi(x_{1}),\phi(x_{2})]\rangle\nonumber\\
&=\Delta_{++}(x_1,x_2)-\Delta_{-+}(x_1,x_2) \nonumber\\
&=\Delta_{+-}(x_1,x_2)-\Delta_{--}(x_1,x_2)\ ,\nonumber\\
\Pi^{R}(x_{1},x_{2})
&=\theta(t_{1}-t_{2})(\Pi^{>}(x_{1},x_{2})-\Pi^{<}(x_{1},x_{2}))\nonumber\\
&=\Pi_{++}(x_1,x_2)-\Pi_{+-}(x_1,x_2) \nonumber\\
&=\Pi_{-+}(x_1,x_2)-\Pi_{--}(x_1,x_2)\ ,\label{selfretarded}\\
\Pi^{A}(x_{1},x_{2})
&=-\theta(t_{2}-t_{1})(\Pi^{>}(x_{1},x_{2})-\Pi^{<}(x_{1},x_{2}))\nonumber\\
&=\Pi_{++}(x_1,x_2)-\Pi_{-+}(x_1,x_2) \nonumber\\
&=\Pi_{+-}(x_1,x_2)-\Pi_{--}(x_1,x_2)\ .\label{selfadvanced}
\end{align}
From Eqs.~(\ref{back1}) and (\ref{forw1}) one obtains the
Kadanoff-Baym equations for the correlation functions 
$\Delta^>$ and $\Delta^<$,
\begin{align}
(\square_{1}+m^{2})\Delta^{>}(x_{1},x_{2})
&=-\int d^{4}x' \left(\Pi^{>}(x_{1},x')\Delta^{A}(x',x_{2})
+ \Pi^{R}(x_{1},x')\Delta^{>}(x',x_{2})\right)\ ,\label{kadanoffbaym1}\\
(\square_{1}+m^{2})\Delta^{<}(x_{1},x_{2})
&=-\int d^{4}x' \left(\Pi^{<}(x_{1},x')\Delta^{A}(x',x_{2})
+\Pi^{R}(x_{1},x')\Delta^{<}(x',x_{2})\right)\ .\label{kadanoffbaym2}
\end{align}

\noindent We now define the real symmetric and antisymmetric correlation functions
\begin{align}
\Delta^{+}(x_{1},x_{2})
&=\frac{1}{2}\langle\{\Phi(x_{1}),\Phi(x_{2})\}\rangle\ ,\label{dplus}\\
\Delta^{-}(x_{1},x_{2})&=i\langle [\Phi(x_{1}),\Phi(x_{2})]\rangle\ ,
\label{dminus}
\end{align}
and self-energies
\begin{align}
\Pi^{+}(x_{1},x_{2})
&=-\frac{i}{2}\left(\Pi^{>}(x_{1},x_{2})+\Pi^{<}(x_{1},x_{2})\right)\ ,
\label{pplus}\\
\Pi^{-}(x_{1},x_{2})
&=\Pi^{>}(x_{1},x_{2})-\Pi^{<}(x_{1},x_{2})\ ,
\label{pminus}
\end{align}
which also determine the retarded and advanced self-energies,
\begin{equation}
\Pi^{R}(x_{1},x_{2})=\theta(t_{1}-t_{2})\Pi^{-}(x_{1},x_{2})\ ,\quad
\Pi^{A}(x_{1},x_{2})=-\theta(t_{2}-t_{1})\Pi^{-}(x_{1},x_{2})\ .
\end{equation}
Adding and subtracting the Kadanoff-Baym equations (\ref{kadanoffbaym1}) and
(\ref{kadanoffbaym2}), one obtains from 
Eqs.~(\ref{retarded})-(\ref{selfadvanced}) and (\ref{dplus})-(\ref{pminus})
an homogeneous equation for $\Delta^-$ and an inhomogeneous equation for
$\Delta^+$,
\begin{eqnarray}
(\square_{1}+m^{2})\Delta^{-}(x_{1},x_{2}) &=& 
-\int d^{3}\textbf{x}'\int_{t_{2}}^{t_{1}} d t' 
\Pi^{-}(x_{1},x')\Delta^{-}(x',x_{2})\;,\label{abzug1}\\
(\square_{1}+m^{2})\Delta^{+}(x_{1},x_{2}) &=&
-\int d^{3}\textbf{x}'\int_{t_{i}}^{t_{1}} dt'
\Pi^{-}(x_{1},x')\Delta^{+}(x',x_{2})\nonumber\\
&& +\int d^{3}\textbf{x}'\int_{t_{i}}^{t_{2}} dt' 
\Pi^{+}(x_{1},x')\Delta^{-}(x',x_{2})\;.\label{addi1}
\end{eqnarray}
We shall refer to these as equations as the first and second Kadanoff-Baym
equation.
$\Delta^{-}$ and $\Delta^{+}$ are known as spectral function and statistical 
propagator (cf.~\cite{ber04}). Together they determine the path ordered
Green's function,
\begin{equation}
\Delta_{C}(x_{1},x_{2}) = \Delta^{+}(x_{1},x_{2})
-\frac{i}{2}\text{sign}_{C}(x^0_{1}-x^0_{2})\Delta^{-}(x_{1},x_{2})\ .
\label{propagatordecomposition}
\end{equation}
$\Delta^{-}$ carries information about the spectrum of the system and 
$\Delta^{+}$ is related to occupation numbers of different modes. 

Using microcausality and the canonical quantization condition for a real scalar
field, 
\begin{align}
[\Phi(x_1),\Phi(x_2)]|_{t_{1}=t_{2}} 
&= [\dot\Phi(x_1),\dot\Phi(x_2)]|_{t_{1}=t_{2}} = 0\ , \\
[\Phi(x_1),\dot\Phi(x_2)]|_{t_{1}=t_{2}} &= i\delta({\bf x}_1-{\bf x}_2)\ ,
\end{align}
one obtains from the definitions (\ref{dplus}) and (\ref{dminus})
\begin{align}
\Delta^{-}(x_{1},x_{2})|_{t_{1}=t_{2}} &= 0\ , \label{con1}\\
\partial_{t_{1}}\Delta^{-}(x_{1},x_{2})|_{t_{1}=t_{2}}
=-\partial_{t_{2}}\Delta^{-}(x_{1},x_{2})|_{t_{1}=t_{2}}
 &= \delta({\bf x}_1-{\bf x}_2)\ , \label{con2}\\
\partial_{t_{1}}\partial_{t_{2}}
\Delta^{-}(x_{1},x_{2})|_{t_{1}=t_{2}} &= 0\ .
\label{con3}
\end{align}

In the following we shall restrict ourselves to systems with spatial 
translational invariance. In this case all two-point functions only depend 
on the difference of spatial coordinates, ${\bf x}_1-{\bf x}_2$, and it is 
convenient to perform a Fourier  transformation. The Green's functions
$\Delta^{\pm}_{\bf q}(t_1,t_2)$ satisfy the two Kadanoff-Baym equations 
\begin{align}
\label{kbe1}
&\Boxq\Delta^{-}_{\q}(t_{1},t_{2})+  
\int_{t_{2}}^{t_{1}} dt'\Pi^{-}_{\q}(t_{1},t')\Delta^{-}_{\q}(t',t_{2})=0\ , \\
&\Boxq\Delta^{+}_{\q}(t_{1},t_{2}) 
+\int_{t_{i}}^{t_{1}} dt'\Pi^{-}_{\q}(t_{1},t')\Delta^{+}_{\q}(t',t_{2})
=\int_{t_{i}}^{t_{2}} dt' \Pi^{+}_{\q}(t_{1},t')\Delta^{-}_{\q}(t',t_{2})\ ,
\label{kbe2}
\end{align}
where $\omega_{\q}^2={\q}^{2}+m^{2}$. The initial conditions (\ref{con1})-(\ref{con3}) for the spectral function become
\begin{align}
\Delta^{-}_{\q}(t_{1},t_{2})|_{t_{1}=t_{2}} &= 0\ , \label{cond1}\\
\partial_{t_{1}}\Delta^{-}_{\q}(t_{1},t_{2})|_{t_{1}=t_{2}}
=-\partial_{t_{2}}\Delta^{-}_{\q}(t_{1},t_{2})|_{t_{1}=t_{2}}
 &= 1\ , \label{cond2}\\
\partial_{t_{1}}\partial_{t_{2}}
\Delta^{-}_{\q}(t_{1},t_{2})|_{t_{1}=t_{2}} &= 0\ .
\label{cond3}
\end{align}

\begin{figure}[t]
  \centering
    \includegraphics[width=12cm]{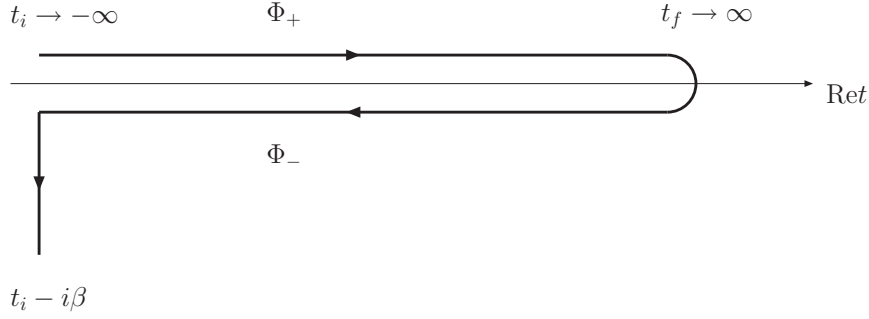}
  \caption{Path in the complex time plane for thermal Green's functions.
\label{contour2}}
\end{figure}

For Green's functions in thermal equilibrium the density matrix in 
Eqs.~(\ref{forw}), (\ref{back}) is $\uprho_{{\rm eq}}=\exp{(-\beta H)}$, where
$H$ is the Hamiltonian of the system, and $\beta=T^{-1}$ is the inverse
temperature. Time coordinates of Green functions now lie on the contour
shown in Figure~2, and one has invariance under time translations so that 
two-point functions only depend on the time difference $t_1-t_2$. After
a Fourier transformation, one obtains the KMS relations \cite{leb96} 
for Green's functions and self-energies,
\begin{align}
\Delta^{+}_{\q}(\omega)&=-\frac{i}{2}\coth\left(\frac{\beta\omega}{2}\right)
\Delta^{-}_{{\q}}(\omega)\ ,\label{DplusKMS}\\
\Pi^{+}_{\q}(\omega)&=-\frac{i}{2}\coth\left(\frac{\beta\omega}{2}\right)
\Pi^{-}_{\q}(\omega)\ .\label{PIplusKMS}
\end{align}

The Kadanoff-Baym equations describe the dynamics of a arbitrary 
nonequilibrium system. Depending on the self-energy and the initial conditions,
the solutions will generally be complicated. An enormous simplification is
achieved for a large medium such that the backreaction of the field 
$\Phi$ can be neglected. Furthermore, we assume that the medium is in thermal equilibrium and, therefore, the self energy of $\Phi$ is time-translation invariant,
\begin{equation}
\Pi_{\q}(t_1,t_2) = \Pi_{\q}(t_1-t_2)\ .
\end{equation}
%
%
In this case also the spectral function is time-translation invariant, as shown
in Appendix~A.1. With these simplifications, the Kadanoff-Baym equations become
\begin{align}
\label{KB1}
\Boxq\Delta^{-}_{\q}(t_{1}-t_{2})&=
-\int_{t_{2}}^{t_{1}} d t' \Pi^{-}_{\q}(t_1-t')\Delta^{-}_{\q}(t'-t_2)\ , \\
\Boxq\Delta^{+}_{\q}(t_{1},t_{2})&=
\int_{t_{i}}^{t_{2}} dt' \Pi^{+}_{\q}(t_{1}-t')\Delta^{-}_{\q}(t'-t_{2})
\nonumber\\
&-\int_{t_{i}}^{t_{1}} dt'\Pi^{-}_{\q}(t_{1}-t')\Delta^{+}_{\q}(t',t_{2})\ .
\label{KB2}
\end{align}
These equations will be solved in Section~4 for general initial conditions.

\section{Stochastic Langevin equation}

Nonequilibrium processes can also be studied by means of Langevin equations
which describe the evolution of the field itself rather than the evolution of 
Green's functions (cf.~\cite{csx84,zin93,yok04,bdh04}). Below we sketch a brief
derivation of the Langevin equation describing a scalar field $\Phi$ coupled
to a large thermal bath with bosonic and fermionic fields $\chi$, following 
the discussion in \cite{bdh04}. We assume that the coupling is of the form
$g\Phi\mathcal{O}[\chi]$ and neglect the backreaction of $\Phi$ on the thermal
bath, which makes the problem solvable. 

The starting point is the nonequilibrium generating functional 
\cite{bdh04,ber04}
\be
{\mathcal Z}[J_+,J_-]=\int D\Phi_{\rm in}^+ D\Phi_{\rm in}^-
\uprho_{\rm in}(\Phi_{\rm in}^+;\Phi_{\rm in}^-)
\int {\mathcal D}\Phi_{\pm}{\mathcal D}\chi_{\beta}
e^{iS[\Phi_{\pm},\chi,J_{\pm}]}\ ,
\label{Z}
\ee 
where the subscript `${\rm in}$' stands for the initial condition. The 
action of the fields $\Phi$ and $\chi$ is given by
\begin{align}\label{action}
S[\Phi_{\pm},\chi,J_{\pm}]=\int^{\infty}_{t_i}&d^4x
\left(\Lm_{\Phi}(\Phi_+) + g\Phi_+{\mathcal O}[\chi_+] + J_+\Phi_+\right. 
\nonumber\\
&\left.-\Lm_{\Phi}(\Phi_-) - g\Phi_-{\mathcal O}[\chi_-] -J_-\Phi_-\right) +
\int_{\Ccnt_{\beta}}d^4x \Lm_{\chi}(\chi) \ ,
\end{align}
where $\Lm_{\Phi}$ is the Lagrangian of a free massive field,
\be
\Lm_{\Phi}={1\over 2}(\partial_{\mu}\Phi)^2-{1\over 2}m^2\Phi^2 \ ,
\ee
and $\uprho_{\rm in}$ stands for the matrix elements of the initial density 
matrix, 
\be
\uprho_{\rm in}(\Phi^+_{\rm in};{\Phi'}^-_{\rm in}) = 
\langle\Phi|\uprho|\Phi'\rangle \ .
\ee
The field $\Phi$ lives on the Keldysh contour $C$ shown in 
Figure~\ref{contour1}. $\Phi_{\pm}(x)$ is the field with the time argument on 
the "forward"($C_+$) and "backward"($C_-$) part of this contour, respectively,
satisfying the boundary conditions
\begin{equation}
\Phi_+(t_i,{\bf x}) = \Phi^+_{\rm in}({\bf x})\ , \quad
\Phi_-(t_i,{\bf x}) = \Phi^-_{\rm in}({\bf x})\ .
\end{equation}
The fields $\chi$ are assumed to be in thermal equilibrium, corresponding to
the contour $\mathcal{C}_{\beta}$ (Figure~\ref{contour2}), which is possible
since the backreaction of $\Phi$ on the thermal bath is neglected. In the
following we shall choose as initial time $t_i=0$.

It is convenient to perform a change of variables in the functional integral
(\ref{Z}),
\begin{align}
&\Psi (x)={1\over 2}\left(\Phi_+(x)+\Phi_-(x)\right)\ ,\\
&R(x)=\Phi_+(x)-\Phi_-(x)\ .
\end{align}
We are  interested in the two-point function of $\Psi$, which couples to the
source term $J=J_+ - J_-$. Integrating out the fields $R$ and $\chi$ one
finds \cite{bdh04},
\begin{align}
{\mathcal Z}[J]=
\int &D\Psi_{\rm in}D\uppi_{\rm in}{\mathcal W}(\Psi_{\rm in};\uppi_{\rm in})
\int{\mathcal D}\Psi{\mathcal D}\xi{\mathcal P}[\xi]
e^{i\int d^4x J(x)\Psi(x)}\nonumber\\
&\times\delta\left[\ddot\Psi_{\q}(t)+\omega_{\q}^2\Psi_{\q}(t)
+\int^t_{0}dt'\Pi^-_{\q}(t-t')\Psi_{\q}(t')-\xi_{\q}(t)\right]\ ;
\label{Z1}
\end{align}
here the measure ${\mathcal P}[\xi]$ is given by
\be\label{weight}
\mathcal{P}[\xi]=\exp\left({1\over 2}\int^{\infty}_0dt
\int^{\infty}_0dt'\xi_{\q}(t)\Pi^+_{\q}(t-t')^{-1}\xi_{-\q}(t')\right]\ ,
\ee
and $\xi_{\q}(t)$ is a {\it stochastic noise}. The Fourier transform 
$\Psi_{\q}(t)$ in (\ref{Z1}) satisfies the initial conditions
\be\label{inipsipi}
\Psi_{\q}(0)=\Psi_{\rm {\q},in}\ , \quad 
\dot\Psi_{\rm {\q},in}(0)=\uppi_{\rm {\q},in}\ .
\ee
The function ${\mathcal W}(\Psi_{\rm in};\uppi_{\rm in})$ is a
functional Wigner transform of the initial density matrix,
\be
{\mathcal W}(\Psi_{\rm in};\uppi_{\rm in})=\int DR_{\rm in}
e^{-\int d^3x\uppi_{\rm in}({\bf x})R_{\rm in}({\bf x})}
\uprho_{\rm in}\left(\Psi_{\rm in}+{R_{\rm in}\over 2};
\Psi_{\rm in}-{R_{\rm in}\over 2}\right)\ .
\ee
For a pure vacuum state $\uprho$ is a product of the two delta functions 
$\delta(\Psi_{\rm in})$ and $\delta(\uppi_{\rm in})$. 

In order to obtain two-point correlators of the field $\Psi$ one has to solve 
the classical {\it stochastic} Langevin equation, 
\begin{equation}\label{langevin}
\left(\partial_t^2 
+\omega_{\q}^2\right)\Psi_{\q}(t)
+\int_{0}^{t}dt' \Pi^{-}_{\q}(t-t')\Psi_{\q}(t')=\xi_{\q}(t)\ ,
\end{equation}
with the initial conditions (\ref{inipsipi}). Since the backreaction of
the field $\Phi$ is neglected, the only relevant correlation functions are
\begin{align}
&\langle\xi_{\q}(t)\rangle=0\ ,\\
&\langle\xi_{\q}(t)\xi_{\q'}(t')\rangle = 
-\Pi^+_{\q}(t-t')\delta(\q+\q')\ .
\label{corr}
\end{align}
The solution of the Langevin equation is conveniently expressed in terms of
an auxiliary function $f_{\q}(t)$ which is defined as solution of the 
homogeneous equation
\begin{equation}\label{auxiliary}
\left(\partial_t^2 +\omega_{\q}^2\right)f_{\q}(t)
+\int_{0}^{t}dt' \Pi^{-}_{\q}(t-t')f_{\q}(t')= 0 \ ,
\end{equation}
with the initial conditions
\begin{equation}
f_{\q}(0) = 0\ ,~~~\dot f_{\q}(0) = 1\ .
\end{equation}
One easily verifies that the solution of the Langevin equation is then given by
\be
\Psi_{\q}(t)=\Psi_{\rm {\q},in}\dot f_{\q}(t)+\Pi_{\rm {\q},in}f_{\q}(t) + 
\int^t_0 dt'f_{\q}(t-t')\xi_{\q}(t')\ .
\label{sol}
\ee

Correlation functions of the scalar field can now be obtained by calculating
the expectation values
\begin{equation}
\langle \Psi_{\textbf{q}_1}(t_1)\ldots \Psi_{\textbf{q}_n}(t_n) \rangle \ ,
\end{equation}
which involve the correlation functions of the stochastic noise and also an
average over the initial conditions. For the simplest case, the two-point 
function, one has
\begin{equation}
\langle \Psi_{\q}(t_1)\Psi_{\q'}(t_2) \rangle 
\equiv g_{\q}(t_1,t_2) \delta({\q}+{\q'})
= g_{\q}(t_2,t_1) \delta({\q}+{\q'})\ .
\end{equation}
From the Langevin equation (\ref{langevin}) one easily derives an analogous
equation for the two-point function,
\begin{align}
&\left(\partial_t^2 +\omega_{\q}^2\right)
\langle\Psi_{\q}(t_{1})\Psi_{\q'}(t_{2})\rangle + 
\int_{0}^{t_1}dt'\Pi_{\q}^{-}(t_{1}-t')
\langle\Psi_{\q}(t')\Psi_{\q'}(t_{2})\rangle \\
&\hspace{2cm}= \langle\xi_{\q}(t_1)\Psi_{\q'}(t_2)\rangle \\
&\hspace{2cm}= \delta({\q}+{\q'})\int^{t_2}_{0}dt'
\Pi^+_{\q}(t_1-t')f_{\q}(t'-t_2)\ ,
\end{align}
which implies
\begin{align}\label{langevin2}
&\left(\partial_t^2 +\omega_{\q}^2\right)g_{\q}(t_1,t_2) + 
\int_{0}^{t_1}dt'\Pi_{\q}^{-}(t_{1}-t')g_{\q}(t',t_2) \\
&\hspace{2cm}= \int^{t_2}_{0}dt'\Pi^+_{\q}(t_1-t')f_{\q}(t'-t_2)\ .
\end{align}

A solution of this equation can be directly obtained from the solution of
the Langevin equation (\ref{langevin}). In the case where the initial
field and its time derivative vanish,
\begin{equation}
\langle \Psi_{{\q},\text{in}} \rangle = 
\langle \dot\Psi_{{\q},\text{in}} \rangle = 0\ ,
\end{equation} 
the relevant averages for the two-point function are
\begin{align}
\langle \Psi_{{\q},\text{in}}\Psi_{{\q},\text{in}} \rangle &=
\delta({\q}+{\q'}) \alpha_{\q}\ ,\\ 
\langle \dot\Psi_{{\q},\text{in}}\dot\Psi_{{\q'},\text{in}} \rangle &= 
\delta({\q}+{\q'}) \beta_{\q}\ ,\\ 
\langle \dot\Psi_{{\q},\text{in}}\dot\Psi_{{\q},\text{in}} \rangle &= 
\delta({\q}+{\q'}) \gamma_{\q}\ .
\end{align} 
Using the solution (\ref{sol}) and the correlations (\ref{corr}) one obtains
the two-point function
\begin{align}
g_{\q}(t_1,t_2) &= \alpha_{\q}\dot f_{\q}(t_1)\dot f_{\q}(t_2) + 
\gamma_{\q}f(t_1)f(t_2) \\
& + \beta_{\q}\left(f_{\q}(t_1)\dot f_{\q}(t_2) 
+ \dot f_{\q}(t_1) f_{\q}(t_2)\right) \\
& + \int^{t_1}_0 dt' \int^{t_2}_0 dt'' 
f_{\q}(t_1-t')\Pi^+_{\q}(t'-t'')f_{\q}(t''-t_2)\ .
\label{solg}
\end{align}

\noindent In the following section we shall see that the auxiliary function $f_{\q}(t)$
and the two-point correlation function $g_{\q}(t_1,t_2)$ are precisely the spectral function and the statistical propagator of the field $\Phi$, 
respectively.

\section{Solving the Kadanoff-Baym equations}\label{solutionsection}

\subsection{The equation for the spectral function}

As proven in Appendix~A.1, the spectral function is time translation invariant,
i.e., it only depends on the time difference $y=t_1-t_2$. Hence, the first 
Kadanoff-Baym equation (\ref{KB1}) takes the form
\begin{equation}
\left(\partial_y^2 + \omega_{\q}^2\right)\Delta_{\q}^-(y) + \int^{y}_{0} dy'
\Pi_{\bf q}^-(y-y')\Delta_{\bf q}^-(y') = 0\ .
\label{eqspectral}
\end{equation}
This equation can be solved by performing a Laplace transformation,
\begin{equation}
\tilde\Delta_{\q}^-(s)=\int^{\infty}_0 dy e^{-sy}\Delta^-_{\q}(y)\ ,
\end{equation}
for which one obtains after a straightforward calculation
\begin{equation}\label{laplacedelta}
\tilde\Delta^-_{\q}(s) = 
\frac{\partial_y\Delta^-_{\q}(0) +  s \Delta^-_{\q}(0)}
{s^2+\omega^2_{\q}+\tilde\Pi^{R}_{\q}(s)}\ ,
\end{equation}
with
\be
\tilde\Pi^{R}_{\q}(s)=\int^{\infty}_0 e^{-s\tm}\Pi^{R}_{\q}(\tm)d\tm 
=\int^{\infty}_0 e^{-s\tm}\Pi^{-}_{\q}(\tm)d\tm=\tilde\Pi^{-}_{\q}(s)\ .
\ee
According to (\ref{laplacedelta}), the general solution of (\ref{eqspectral}) 
depends on two parameters,
the values of $\Delta^-_{\q}$ and $\partial_y\Delta^-_{\q}$ at $y=0$. 
Using the inverse Laplace transform one finds
\begin{equation}
\Delta^-_{\q}(y) = \left(\partial_y\Delta^-_{\q}(0)
+ \Delta^-_{\q}(0)\partial_y\right) 
\int_{\Ccnt_B}{ds\over 2\pi i}
{e^{s\tm}\over s^2+\omega^2_{\q}+\tilde\Pi^{-}_{\q}(s)}\ .
\label{general}
\end{equation}
Here $\Ccnt_B$ is the Bromwich contour (see Figure~3): The part parallel to the 
imaginary axis is chosen such that all singularities of the integrand are to 
its left; the second part is the semicircle at infinity which closes the 
contour at ${\rm Re}(s)<0$. Since the integrand  of (\ref{general}) has 
singularities only on the imaginary axis, the second part can be deformed to 
run parallel to the imaginary axis as well: 
$\Ccnt_B\to \int^{i\infty+\e}_{-i\infty+\e}+\int^{-i\infty-\e}_{i\infty-\e}$. 

The spectral function $\Delta^-_{\q}(y)$ satisfies the boundary conditions 
(\ref{cond1}) and (\ref{cond2}), which implies 
\begin{equation}
\Delta^-_{\q}(y) = \int_{\Ccnt_B}{ds\over 2\pi i}
{e^{s\tm}\over s^2+\omega^2_{\q}+\tilde\Pi^{-}_{\q}(s)}\ .
\label{brom}
\end{equation}
This result can be further simplified by making use of the analytic properties 
of the self-energy $\tilde\Pi^{-}(s)$. On the real axis $\tilde\Pi^{-}(s)$ is 
real, while on the parts of the contour which are parallel to the imaginary 
axis one has 
\be
\tilde\Pi^{-}(i\omega\pm\e) = {\rm Re}\Pi^R_{\q}(\omega)\pm 
i{\rm Im}\Pi^R_{\q}(\omega)\ ,
\ee
with
\begin{equation}
{\rm Im} \Pi^{R}_{\bf{q}}(\omega)=\frac{1}{2i}
\left(\Pi^{R}_{\bf{q}}(\omega+i\epsilon)
-\Pi^{R}_{\bf{q}}(\omega-i\epsilon)\right)\ .
\end{equation}
Hence, the expression (\ref{brom}) takes the form
\be
\Delta^-_{\q}(\tm)=i\int^{\infty}_{-\infty}\frac{d\omega}{2\pi} 
e^{-i\omega \tm}\rho_{\q}(\omega)\ ,
\label{dmin}
\ee
where the spectral function $\rho_{\q}(\omega)$ is given in terms of real and 
imaginary part of the self-energy $\Pi^R_{\q}(\omega)$,  
\be
\rho_{\q}(\omega)={-2{\rm Im}\Pi^R_{\q}(\omega)+2\omega\epsilon\over 
[\omega^2-\omega_{\q}^2-{\rm Re}\Pi^R_{\q}(\omega)]^2+
[{\rm Im}\Pi^R_{\q}(\omega)+\omega\epsilon]^2} =
i\tilde\Delta_{\q}^-(i\omega)\ . \label{spectralfunction}
\ee
Note that $\im\Pi^R_{\q}(\omega)$ and $\re\Pi^R_{\q}(\omega)$ are odd and even
functions, respectively, which implies that $\Delta_{\q}(y)$ is 
real. Further properties of this solution are discussed in Appendix~A. 
Let us recall that the expression (\ref{spectralfunction}) is obtained after
neglecting the backreaction of the field $\Phi$ on the thermal bath. This is 
the reason why the self-energy and the spectral function are time translation 
invariant.

\begin{figure}[t]
\centering
\includegraphics[width=10 cm]{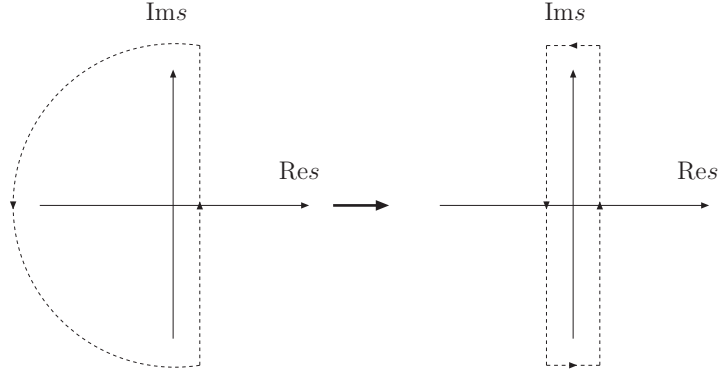}
\caption{Bromwich contour}
\label{Bromwich}
\end{figure}

The self-energy $\Pi^R_{\q}(\omega)$, and consequently the spectral function
$\rho_{\q}(\omega)$, are divergent and have to be renormalized. This can be 
done by the usual mass and wave function renormalization at zero temperature. 
In (\ref{spectralfunction}) $\omega_{\q}^2$ is replaced by  
$\omega^2_{\textbf{q} (0)} = m_0^2 + \textbf{q}^2$, where $m_0$ is the bare mass of the field
$\Phi$. The difference between bare and renormalized mass squared is determined
by requiring that at zero temperature the spectral function has a pole at 
$\omega_{\q}^2 = m^2 + \textbf{q}^2$,
\begin{equation}\label{polgleichungT0}
\omega_{\q}^{2}- \omega^2_{\textbf{q} (0)} - 
{\rm Re}\Pi_{\q}^R(\omega_{\q})|_{T=0} = 0\ .
\end{equation}
Expanding the self-energy around around $\omega_{\q}$, a further divergence
can be absorbed in a wave function renormalization constant,
\begin{equation}
{\rm Re} \Pi_{\q}^R(\omega) = 
{\rm Re} \Pi_{\q}^R(\omega_{\q})|_{T=0} 
+ \left(1 - Z^{-1}\right)\left(\omega^2 - \omega_{\q}^2\right)
+ {\rm Re} \hat{\Pi}_{\q}^R(\omega)\ ,
\end{equation}
where ${\rm Re} \hat{\Pi}_{\q}^R(\omega)$ is the finite part and
\begin{equation}
Z^{-1} = 1-\frac{1}{2\omega_{\q}}\frac{\partial 
{\rm Re}\Pi^{R}_{\q}(\omega)}{\partial\omega}\Big|_{\omega=\omega_{\q}, T=0}\ .
\end{equation}
The spectral function (\ref{spectralfunction}) now takes the form
\begin{equation}
\rho_{\q}(\omega) = 
Z\frac{-2Z{\rm Im}\Pi^{R}_{\q}(\omega)+2\omega\epsilon}
{\left(\omega^{2}-\omega_{\q}^{2}-Z{\rm Re}\hat{\Pi}_{\q}^R(\omega)\right)^2
+\left(Z{\rm Im}\Pi^{R}_{\q}(\omega)+\omega\epsilon\right)^{2}}\ .
\end{equation}
Introducing the renormalized field operator $\Phi_{r}=\sqrt{Z}\Phi$, one 
obtains the renormalized spectral function 
$\rho_{\q}^{r}(\omega) = Z \rho_{\q}(\omega)$ 
in terms of the renormalized self-energy 
$\Pi^{R,r}_{\q}(\omega) = Z\hat{\Pi}^{R}_{\q}(\omega)$,
\begin{equation}\label{spectralfunctionReno}
\rho^{r}_{\q}(\omega) = 
\frac{-2 {\rm Im}\Pi^{R,r}_{\q}(\omega)+2\omega\epsilon}
{\left(\omega^{2}-\omega_{\q}^{2}-{\rm Re}\Pi_{\q}^{R,r}(\omega)\right)^2
+\left({\rm Im}\Pi^{R,r}_{\q}(\omega)+\omega\epsilon\right)^{2}}\ .
\end{equation}
The divergencies of spectral function and statistical propagator can be
removed in the same way by mass and wave function renormalization at zero
temperature. In the following we shall drop the superscript `$r$' to keep the
notation simple.

The spectral function describes a quasi-particle resonance at finite 
temperature with energy $\Omega_{\q}$,
\begin{equation}
\Omega^2_{\q}-\omega_{\q}^{2}-{\rm Re}\Pi^{R}_{\q}(\Omega_{\q}) = 0,\quad
\Omega^2_{\q}|_{T=0} = \omega^2_{\q}\ ,
\end{equation}
and decay width
\begin{equation}
\Gamma_{\q} \simeq -\frac{1}{\Omega_{\q}}\text{Im}\Pi^{R}_{\q}(\Omega_{\q})\ .
\end{equation}
For simplicity, we have neglected the effect of $\text{Im}\Pi^{R}_{\q}$ 
on the quasi-particle energy. The correction $\delta\Omega_{\q} = 
{\cal O}(\Gamma_{\q}^2)$ is evaluated in Section~6. 

In a free theory $\im\Pi^{R}_{\q}(\omega) = 0$, and
(\ref{spectralfunctionReno}) is a representation of the $\delta$-function. 
The spectral function (\ref{dmin}) then oscillates without damping, i.e.,
there are no dissipative effects. 
Dissipation arises either from $\Phi$ decays and inverse decays or, 
similar to Landau damping, from scattering processes with particles in the 
plasma. Which of these mechanisms dominates the dissipative effects and 
therefore the equilibration process depends on the position of the 
quasi-particle pole relative to the masses of particles in the thermal bath.
A specific example will be discussed in Section~7. For small width the 
spectral function is well approximated by the Breit-Wigner function. The
relevant formulae are collected in Appendix~A.5.

\subsection{Solution for the statistical propagator}

We are now ready to solve the second Kadanoff-Baym equation (\ref{KB2}) for 
the statistical propagator, which for initial time $t_i=0$ is given by
\begin{equation}
\Boxq\Delta_{\bf q}^+(t_1,t_2)+\int^{t_1}_{0} dt'
\Pi_{\bf q}^-(t_1-t')\Delta_{\bf q}^+(t',t_2)=\zeta(t_1,t_2)\label{kb2}\ ,
\end{equation}
with
\begin{equation}
\zeta(t_1,t_2)=\int^{t_2}_{0} dt'
\Pi_{\q}^+(t_1-t')\Delta_{\bf q}^-(t'-t_2)\ .
\end{equation}
One easily verifies that the solution can be expressed as
\begin{equation}
\Delta^+_{\q}(t_1,t_2)=\hat{\Delta}_{\q}^+(t_1,t_2) 
+\int_0^{t_1}dt'\Dm (t_1-t')\zeta(t',t_2)\ ,
\label{sol2}
\end{equation}
where $\hat{\Delta}_{\q}^+(t_1,t_2)$ satisfies the homogeneous equation
\begin{equation}
\Boxq\hat{\Delta}_{\q}^+(t_1,t_2)+\int^{t_1}_{0} dt'
\Pi_{\q}^-(t_1-t')\hat{\Delta}_{\q}^+(t',t_2) = 0\ .
\end{equation}

The homogeneous equation is identical to (\ref{eqspectral}), with $t_2$ 
playing the role of a parameter. We can therefore read off the general 
solution from (\ref{general}),
\be
\hat{\Delta}_{\q}^+(t_1,t_2)=A_{\q}(t_2)\dot\Delta_{\q}^-(t_1) 
+ B_{\q}(t_2)\Dm(t_1)\ . 
\label{free}
\ee
Using the symmetry 
$\hat{\Delta}_{\q}^+(t_1,t_2) = \hat{\Delta}_{\q}^+(t_2,t_1)$, one obtains
\be\label{solab}
A_{\q}(t_2)\dot\Delta_{\q}^-(t_1)+B_{\q}(t_2)\Dm(t_1) = 
A_{\q}(t_1)\dot\Delta_{\q}^-(t_2)+B_{\q}(t_1)\Dm(t_2)\ .
\ee 
Together with the boundary conditions (\ref{cond1})-(\ref{cond3}), 
$\Delta_{\q}^-(0) = \ddot\Delta_{\q}^-(0) = 0$ and $\dot\Delta_{\q}^-(0) = 1$,
this implies
\be\label{abf}
A_{\q}(t) = A_{\q}(0)\dot\Delta_{\q}^-(t) + B_{\q}(0) \Dm(t)\ ,\quad
B_{\q}(t) = \dot A_{\q}(0)\dot\Delta_{\q}^-(t) + \dot B_{\q}(0) \Dm(t)\ .
\ee
Inserting $A_{\q}(t)$ and $B_{\q}(t)$ in (\ref{solab}) and using the symmetry
of $\hat{\Delta}_{\q}^+(t_1,t_2)$, one finds $B_{\q}(0)=\dot A_{\q}(0)$.
The initial state of the system is therefore characterized by three constants,
which can be chosen as
\begin{eqnarray}
\Delta^{+}_{\q,\text{in}} &=& 
\Delta^{+}_{\q}(t_{1},t_{2})|_{t_{1}=t_{2}=0} = A_{\q}(0)\ ,\label{in1}\\
\dot{\Delta}^{+}_{\q,\text{in}} &=& 
\partial_{t_{1}}\Delta^{+}_{\q}(t_{1},t_{2})|_{t_{1}=t_{2}=0} 
= \partial_{t_{2}}\Delta^{+}_{\q}(t_{1},t_{2})|_{t_{1}=t_{2}=0}
= B_{\q}(0)=\dot A_{\q}(0)\ , \label{in2}\\
\ddot{\Delta}^{+}_{\q,\text{in}} &=& 
\partial_{t_{1}}\partial_{t_{2}}
\Delta^{+}_{\q}(t_{1},t_{2})|_{t_{1}=t_{2}=0} = \dot B_{\q}(0)\ .\label{in3}
\end{eqnarray}

From Eqs.~(\ref{sol2}), (\ref{free}), (\ref{abf}) and the initial conditions
(\ref{in1})-(\ref{in3}) we now obtain the full solution for the statistical
propagator,
\begin{eqnarray}\label{solution}
\Delta^{+}_{\q}(t_{1},t_{2}) &=& 
\Delta^{+}_{\q,\text{in}}
\dot{\Delta}^{-}_{\q}(t_{1})\dot{\Delta}^{-}_{\q}(t_{2})
+\ddot{\Delta}^{+}_{\q,\text{in}}
\Delta^{-}_{\q}(t_{1})\Delta^{-}_{\q}(t_{2})\nonumber\\
&+&\dot{\Delta}^{+}_{\q;\text{in}}
\left(\dot{\Delta}^{-}_{\q}(t_{1})\Delta^{-}_{\q}(t_{2})
+\Delta^{-}_{\q}(t_{1})\dot{\Delta}^{-}_{\q}(t_{2})\right)\nonumber\\
&+& \Delta^{+}_{\q,\text{mem}}(t_{1},t_{2})\ ,
\end{eqnarray}
where
\begin{equation}
\Delta^{+}_{\q,\text{mem}}(t_{1},t_{2}) =
\int_{0}^{t_{1}}dt'\int_{0}^{t_{2}}dt''
\Delta^{-}_{\q}(t_{1}-t')\Pi^{+}_{\q}(t'-t'')\Delta^{-}_{\q}(t''-t_{2})\ .
\end{equation}
This contribution to the statistical propagator, which is independent of the 
initial conditions, is often referred to as {\it memory integral}. It can be 
expressed in the form
\be
\Delta^{+}_{\q,\text{mem}}(t_{1},t_{2}) =
-\int^{\infty}_{-\infty} \frac{d\omega}{2\pi} e^{-i\omega(t_{1}-t_{2})}
{\mathcal H}_{\q}^*(t_1,\omega){\mathcal H}_{\q}(t_2,\omega)
\Pi^{+}_{\q}(\omega)\ ,
\label{memint}
\ee
where \cite{bdh04}
\be
{\mathcal H}_{\q}(t,\omega)=\int^{t}_0d\tau e^{-i\omega\tau}\Dm(\tau)\ .
\ee
The expression (\ref{memint}) will be the basis of our numerical analysis 
in Section~7.

\section{Thermal equilibrium and quasi-particles}

Let us now verify that the solution (\ref{solution}) for the statistical
propagator approaches thermal equilibrium at late times. This means that
the quantity
\begin{equation}
\Delta^+_{\q}(t,\omega) = \int_{-2t}^{2t}dy e^{i\omega y}
\Delta^+_{\q}\left(t+\frac{y}{2},t-\frac{y}{2}\right)\ ,
\end{equation}
which becomes a Fourier transform for $t\rightarrow \infty$, satisfies the 
KMS condition asymptotically,
\begin{equation}\label{DeltaPlusKMS}
\Delta^{+}_{\q}(\infty,\omega) =
-\frac{i}{2}\coth\left(\frac{\beta\omega}{2}\right)\Delta^{-}_{\q}(\omega)\ .
\end{equation}

For late times only the memory integral is relevant, since  
$\Delta^{-}_{\q}(t)$ and $\dot\Delta^{-}_{\q}(t)$ fall off exponentially
for $\tp\gg 1/\Gamma$. One then obtains
\begin{equation}\label{larget}
\Delta^{+}_{\q}(\infty,\omega) = 
\Delta^{+}_{\q,\text{mem}}(\infty,\omega) 
= -|{\mathcal H}_{\q}(\infty,\omega)|^2\Pi^{+}_{\q}(\omega)\ .
\end{equation}
The quantity ${\mathcal H}_{\q}(\infty,\omega)$ is the Laplace
transform of the spectral function,
\begin{align}
{\mathcal H}_{\q}(\infty,\omega)&=
\int^{\infty}_0 d\tau e^{-i(\omega-i\epsilon)\tau}\Dm(\tau)\nonumber\\
&= \tilde{\Delta}^-_{\q}(i\omega+\epsilon)\nonumber\\
&={1\over s^2+\omega_q^2+\tilde\Pi_{\q}(s)}\Big|_{s=i\omega+\epsilon}
\nonumber\\
&=-\frac{1}{\omega^2-\omega_q^2-\text{Re}\Pi_{\q}^R(\omega)
-i\text{Im}\Pi^R_{\q}(\omega)}\ ,
\end{align}
which yields
\begin{align}\label{htinfty}
|{\mathcal H}_{\q}(\infty,\omega)|^2 &= 
\frac{1}{(\omega^2-\omega_{\q}^2-{\rm Re}\Pi^R_{\q}(\omega))^2 
+ ({\rm Im}\Pi^R_{\q}(\omega))^2} \nonumber\\
&= -\frac{\rho_{\q}(\omega)}{2\im\Pi^R_{\q}(\omega)}\ .
\end{align}
Inserting this expression into (\ref{larget}), using the KMS condition 
for the self-energy and (\ref{ipa}),
\begin{equation}
\Pi^{-}_{\q}(\omega)=2i{\rm Im}\Pi^{R}_{\q}(\omega)\ , \nonumber
\end{equation}
one obtains (cf.~(\ref{dmin}),(\ref{spectralfunction})), 
\begin{align}
\Delta^{+}_{\q}(\infty,\omega)
&= -\coth\left(\frac{\beta\omega}{2}\right)
\frac{{\rm Im}\Pi^{R}_{\q}(\omega)}
{(\omega^2-\omega_{\q}^2-{\rm Re}\Pi^R_{\q}(\omega))^2 
+ ({\rm Im}\Pi^R_{\q}(\omega))^2}
\nonumber\\
&= -\frac{i}{2}\coth\left(\frac{\beta\omega}{2}\right)
\Delta^{-}_{\q}(\omega)\ .
\end{align}
Hence, our solution for the statistical propagator indeed fulfills
the KMS condition (\ref{DplusKMS}) in the limit $t\rightarrow\infty$, which 
proves that the system reaches thermal equilibrium. For a specific example 
the approach to equilibrium will be studied numerically in Section~7.

It is instructive to evaluate the statistical propagator in thermal equilibrium
at equal times, i.e., $y=t_1-t_2=0$,
\begin{equation}\label{DeltapT}
\Delta^+_{\q}\big|_{y=0} = 
\frac{1}{2}\int^{\infty}_{-\infty}\frac{d \omega}{2\pi}  
\coth\left({\beta\omega\over 2}\right)\rho_{\q}(\omega)\ .
\end{equation}
For a free field one has
\begin{equation}\label{Deltapfree}
\rho_{\q}(\omega) = 
2\pi\text{sign}(\omega)\delta(\omega^2-\omega_{\q}^2)\ ,
\end{equation}
which yields the well know result
\begin{equation}\label{delta+free}
\Delta^+_{\q}|_{y=0} = 
\frac{1}{\omega_{\q}}\left(\frac{1}{2} + n_\text{B}(\omega_{\q})\right)\ ,
\end{equation}
with the temperature dependent Bose-Einstein distribution function
\begin{equation}
n_\text{B}(\omega_{\q}) = \frac{1}{e^{\beta\omega_{\q}}-1}\ .
\end{equation}

Generically, the interaction with the thermal bath changes the energy 
$\omega_{\q}$ of a free particle to a temperature dependent complex
energy $\hat{\Omega}_{\q}$ which appears as a pole of the spectral 
function $\rho_{\q}(\omega)$ and the integrand of (\ref{DeltapT}).
The spectral function then has two poles in the upper plane,  $\hat{\Omega}_{\q}$
and  $-\hat{\Omega}_{\q}^*$, which are determined by the condition
\begin{equation}
\hat{\Omega}_{\q} - \left(\omega_{\q}^2 + 
\Pi^R_{\q}\left(\hat{\Omega}_{\q}\right)\right)^{1/2} = 0\ .
\end{equation}
Assuming that the integral can be closed in the upper half-plane,
one obtains for the statistical propagator in equilibrium,\footnote{Here we restrict ourselves to the case where there are no additional poles.}
\begin{equation}\label{delta+quasi}
\Delta^+_{\q}|_{y=0} = \text{Re}\left(\frac{1}{\hat{\Omega}_{\q}}
\left(\frac{1}{2} + n_\text{B}(\hat{\Omega}_{\q})\right)
\right)\ .
\end{equation}
Compared to (\ref{Deltapfree}), the Bose-Einstein distribution function
has been replaced by the complex distribution function 
$n_{\text{B}}(\hat{\Omega}_{\q})$. 

At high temperatures, where $\beta\omega_{\q} \ll 1$, the Bose-Einstein
distribution has a well-known infrared divergence,
\begin{equation}
n_{\text{B}}(\omega_{\q}) \simeq
\frac{1}{\beta\omega_{\q}} \gg 1 \ .
\end{equation}
For quasi-particles, where $\omega_{\q}$ is replaced by 
$\hat{\Omega}_{\q} = \Omega_{\q} + i \Gamma_{\q}/2$,
this divergence is cut off by the finite width,
\begin{equation}
|n_{\text{B}}(\hat{\Omega}_{\q})| \simeq 
\frac{1}{|\beta(\Omega_{\q} + \frac{i}{2}\Gamma_{\q})|}
\leq \frac{2}{\beta\Gamma_{\q}}\ , 
\end{equation}
which remains finite even if the real part $\Omega_{\q}$ vanishes.

Comparison of equations (\ref{delta+free}) and (\ref{delta+quasi}) suggests
that in thermal equilibrium the $\Phi$ particles may form a gas of 
quasi-particles. This question can be clarified by evaluating energy density
and pressure of the $\Phi$ particles. Since the expectation value of $\Phi$
vanishes, one obtains from the energy momentum tensor\footnote{We use the
convention $\text{diag}\ (\eta_{\mu\nu})=(1,-1,-1,-1).$}
\begin{equation}
T_{\mu\nu} = \partial_{\mu}\Phi\partial_{\nu}\Phi - \eta_{\mu\nu} L\ 
\end{equation}
for the contribution of a mode with momentum $\q$ to energy density and 
pressure,
\begin{align}
&\epsilon_{\q} = \langle T_{00} \rangle|_{\q} = 
\frac{1}{2}\langle\dot\Phi^2 + 
(\vec{\nabla}\Phi)^2 + m^2 \Phi^2\rangle|_{\q} \ ,\\
&p_{\q} = \langle T_{ii} \rangle|_{\q} = 
\langle \frac{1}{3}({\bf \nabla}\Phi)^2 
+ \frac{1}{2}(\dot\Phi^2 - ({\bf \nabla}\Phi)^2 - m^2 \Phi^2)\rangle|_{\q}\ .
\end{align}
This yields for the energy density
\begin{align}
\epsilon_{\q}(\infty) &= \frac{1}{2}\left(\partial_{t_1}\partial_{t_2} 
+ \omega_{\q}^2 \right)\Delta^+_{\q}(t_1,t_2)\big|_{t_1=t_2=\infty} \nonumber\\
&= \frac{1}{2}\left(\Omega_{\q}^2 + \omega_{\q}^2\right)\frac{1}{\Omega_{\q}}
\left(\frac{1}{2} + n_\text{B}(\Omega_{\q})\right)\ ,
\end{align}
and for the pressure
\begin{align}
p_{\q}(\infty) &= \left(\frac{1}{3}\textbf{q}^2 
+ \frac{1}{2}\left(\partial_{t_1}\partial_{t_2} 
- \omega_{\textbf{q}}^2 \right)\right)\Delta^+_{\q}(t_1,t_2)\big|_{t_1=t_2=\infty} \nonumber\\
&= \left(\frac{1}{3}\textbf{q}^2 
+ \frac{1}{2}\left(\Omega_{\textbf{q}}^2 - \omega_{\textbf{q}}^2\right)\right)
\frac{1}{\Omega_{\textbf{q}}}
\left(\frac{1}{2} + n_\text{B}(\Omega_{\q})\right)\ ,
\end{align}
where, for simplicity, we have neglected the quasi-particle width.

In summary, the energy momentum tensor in thermal equilibrium can be expressed
as sum of a quasi-particle gas contribution and a temperature dependent
`vacuum' term,
\begin{equation}
\langle T_{\mu\nu}\rangle|_{\q} = 
u_{\mu}u_{\nu}\left(\epsilon^{\text{QP}}_{\q} + p^{\text{QP}}_{\q}\right) 
- \eta_{\mu\nu} p^{\text{QP}}_{\q} + \eta_{\mu\nu} \kappa^{\text{VAC}}_{\q}\ .
\end{equation}
Here $u^{\mu} = (1,\vec{0})$ is the 4-velocity of the thermal bath, and
\begin{align}
\epsilon^{\text{QP}}_{\q} &= 
\Omega_{\q}\left(\frac{1}{2} + n_\text{B}(\Omega_{\q})\right)\ ,\\
p^{\text{QP}}_{\q} &= \frac{1}{3}\frac{\textbf{q}^2}{\Omega_{\q}}
\left(\frac{1}{2} + n_\text{B}(\Omega_{\q})\right)\ ,\\
\kappa^{\text{VAC}}_{\q} &= 
\frac{\omega_{\q}^2 - \Omega_{\q}^2}{2\Omega_{\q}}
\left(\frac{1}{2} + n_\text{B}(\Omega_{\q})\right)\ .
\end{align}
Energy density and pressure of the quasi-particle gas agree with the 
corresponding expressions for a free gas, with the energy $\omega_{\q}$ of
a free particle replaced by the quasi-particle energy $\Omega_{\q}$. The
`vacuum contribution' $\kappa^{\text{VAC}}_{\q}$ vanishes for  
$\Omega_{\q} = \omega_{\q}$. For large thermal effects, i.e. 
$\Omega_{\q} \gg \omega_{\q}$ or $\Omega_{\q} \ll \omega_{\q}$, the equation 
of state differs significantly from the one of a free gas. Note that for 
$\Omega_{\q}^2 < \omega_{\q}^2$,
the pressure can even become negative!

\section{Comparison with Boltzmann equations}

The time evolution of nonequilibrium systems is usually studied by means of
Boltzmann equations for particle number densities. However, this notion does
not have a well defined physical meaning in a nonequilibrium process. For
a dilute, weakly coupled gas the number density of `free particles' may be
a good approximation, and in some cases the the effect of a medium can be
taken into account by considering quasi-particles. In general, however, one has to study the
time evolution of Green's functions, in
particular if quantum interferences are important.

In order to determine the range of validity of the Boltzmann approximation, we 
shall consider in this section the time evolution of an observable, the energy 
density. The exact expression can be obtained from the statistical propagator, 
and approximations are given by solutions of Boltzmann equations. In this way, 
the description of the nonequilibrium process by means of Green's functions on
the one hand, and Boltzmann equations on the other hand, can be directly
compared, based on the same observable.

Consider the Boltzmann equation for a dilute gas of $\Phi$ particles. 
The competition between a gain and a loss term determines the change of the
particle number density \cite{wel83},
\begin{equation}\label{boltz1}
\partial_{t}n_{\q}(t)=(1+n_{\q}(t))\gamma^{<}_{\q}-n_{\q}(t)\gamma^{>}_{\q}\ ,
\end{equation}
where production and decay rates satisfy the KMS relation and are obtained 
from the self-energy of the field $\Phi$, 
\begin{align}
\gamma^{>}_{\q} &= e^{-\beta\omega_{\q}}\gamma^{<}_{\q}
\equiv n_{\text{B}}(\omega_{\q}) \gamma_{\q} \ ,\\
\gamma_{\q} &= -\frac{{\rm Im}\Pi^R_{\q}(\omega_{\q})}{\omega_{\q}}\ .
\end{align}
Using these relations, the Boltzmann equation (\ref{boltz1}) can be 
written in the form
\begin{equation}
\partial_{t}n_{\q}(t) = -\gamma_{\q}(n_{\q}(t)-n_{\text{B}}(\omega_{\q}))\ ,
\end{equation}
with the obvious solution
\begin{equation}\label{BoltzmannSol}
n_{\q}(t)=n_{\text{B}}(\omega_{\q})+\left(n_{\q}(0)
-n_{\text{B}}(\omega_{\q})\right)e^{-\gamma_{\q}t}\ .
\end{equation}

For comparison with the Kadanoff-Baym equations we now consider instead of the 
number density the energy density of a mode with momentum $\q$, normalized
to the energy of a single quantum,
\begin{equation}
\hat{\epsilon}_{\q}(t) \equiv \frac{\epsilon_{\q}(t)}{\omega_{\q}}
= \frac{1}{2} + n_{\q}(t)\ .
\end{equation}
The deviation from the equilibrium density,
\begin{equation}
\hat{\epsilon}_{\q}(t) = 
\hat{\epsilon}^{\text{free}}_{\q} + 
\delta\hat{\epsilon}_{\q}(t)\ ,
\end{equation}
with
\begin{equation} 
\hat{\epsilon}_{\q}(\infty) \equiv 
\hat{\epsilon}^{\text{free}}_{\q} = 
\frac{1}{2} + n_{\text{B}}(\omega_{\q})\ ,
\end{equation}
satisfies the differential equation
\begin{equation}
(\partial_t + \gamma_{\q})\delta\hat{\epsilon}_{\q}(t) = 0\ .
\end{equation} 

The modification of the spectral function in a thermal bath 
(cf.~(\ref{spectralfunction}))
suggests to replace the equilibrium value and the evolution equation for the
energy density by the expressions
\begin{equation}
\hat{\epsilon}_{\q}(t) = 
\hat{\epsilon}^{\text{QP}}_{\q} + 
\delta\hat{\epsilon}_{\q}(t)\ ,\quad 
\hat{\epsilon}^{\text{QP}}_{\q} = 
\frac{1}{2} + n_{\text{B}}(\Omega_{\q})\ , 
\end{equation}
and
\begin{equation}\label{qpboltz}
(\partial_t + \Gamma_{\q})\delta\hat{\epsilon}_{\q}(t) = 0\ ,
\end{equation} 
where the quasi-particle width is given by (cf.~Appendix A.5)
\begin{equation}\label{qpwidth}
\Gamma_{\q}=-Z_{\q}\frac{{\rm Im}\Pi^{R}_{\q}(\Omega_{\q})}{\Omega_{\q}}\ ,
\quad
Z_{\q}^{-1}=1 - \frac{1}{2\Omega_{\q}}\frac{\partial}{\partial \omega}
\text{Re}\Pi^R_{\q}(\omega)\big|_{\Omega_{\q}}\ .
\end{equation}
As long as the interaction of the field $\Phi$ with the thermal bath can
be treated perturbatively, the difference between solutions of the two
Boltzmann equations for particles and quasi-particles, respectively,
should be small. When the quasi-particle width becomes large, however,
the use of first-order differential equations, which are local in time,
becomes clearly questionable.

As discussed in Section~5, the exact time dependence of the energy density 
can be directly obtained from the statistical propagator,
\begin{equation}\label{energy}
\hat{\epsilon}_{\q}(t) = \frac{1}{2\omega_{\q}}
\left(\partial_{t_1}\partial_{t_2} 
+ \omega_{\q}^2 \right)\Delta^+_{\q}(t_1,t_2)\big|_{t_1=t_2=t}\ , \nonumber
\end{equation}
which satisfies the Kadanoff-Baym equation (\ref{kadanoffbaym2}),
\begin{equation}
\Boxq\Delta^{+}_{\q}(t_{1},t_{2}) +
\int_{0}^{t_{1}} dt'\Pi^{-}_{\q}(t_1-t')\Delta^{+}_{\q}(t',t_2)
=\int_{0}^{t_{2}} dt' \Pi^{+}_{\q}(t_1-t')\Delta^{-}_{\q}(t'-t_2)\ .
\end{equation}
For large times, $t\gg 1/\Gamma_{\q}$, the dependence on the initial values
at $t_i =0$ can be neglected, and one obtains
\begin{equation}
\Boxq\Delta^{+}_{\q}(t_{1},t_{2}) +
\int_{-\infty}^{\infty} dt'\left(\Pi^R_{\q}(t_1-t')\Delta^{+}_{\q}(t',t_2)
+ i\Pi^{+}_{\q}(t_1-t')\Delta^A_{\q}(t'-t_2)\right) = 0\ .
\end{equation}
Changing time variables,
\begin{equation}
t = \frac{t_1+t_2}{2}\ , \quad y = t_1-t_2\ , \quad
\Delta^{+}_{\q}\left(t;y\right) \equiv \Delta^{+}_{\q}(t_1,t_2)\ ,
\end{equation}
and expanding,
\begin{equation}
\Delta^{+}_{\q}\left(\frac{t'+t_2}{2};t'-t_2\right) =  
\Delta^{+}_{\q}\left(t;t'-t_2\right) +
\frac{t'-t_1}{2}\partial_{t}
\Delta^{+}_{\q}\left(t;t'-t_2\right) + \ldots\ ,
\end{equation}
one finds for the Fourier transforms with respect to the time differences, 
\begin{eqnarray}\label{DerExpStartGleichung}
\lefteqn{\left(\frac{1}{4}\partial_t^2-i\omega\partial_t 
- \omega^2 + \omega_{\q}^2\right)\Delta^+_{\q}(t;\omega)}\nonumber\\
&=&-\Pi^R_{\q}(\omega)\Delta^{+}_{\q}(t;\omega)
-i\Pi^{+}_{\q}(\omega)\Delta^{A}_{\q}(t;\omega)
- \frac{i}{2}\frac{\partial\Pi^R_{\q}(\omega)}{\partial\omega}
\frac{\partial\Delta^+_{\q}(t;\omega)}{\partial t}\ .
\end{eqnarray}

Using the relations (\ref{dm}) - (\ref{par}), one obtains from the real and 
the imaginary part
of this complex equation two equations for the real quantity 
$\Delta^+_{\q}(t,\omega)$,
\begin{align}
\left(\frac{1}{4}\partial_t^2 - \omega^2 + \omega_{\q}^2\right)
\Delta^+_{\q}(t,\omega) &=
-\text{Re}\Pi^R_{\q}(\omega)\Delta^{+}_{\q}(t,\omega)
+\Pi^{+}_{\q}(\omega)\text{Im}\Delta^{A}_{\q}(t,\omega)\nonumber\\
&\quad +\frac{1}{2}\frac{\partial\text{Im}\Pi^R_{\q}(\omega)}{\partial\omega}
\frac{\partial\Delta^+_{\q}(t,\omega)}{\partial t} +\ldots\ ,\label{reeq}\\
\omega \frac{\partial}{\partial t}\Delta^{+}_{\q}(t,\omega)&= 
\text{Im}\Pi^R_{\q}(\omega)\Delta^{+}_{\q}(t,\omega)
+\Pi^{+}_{\q}(\omega)\text{Re}\Delta^{A}_{\q}(t,\omega)\nonumber\\
&\quad +\frac{1}{2}\frac{\partial\text{Re}\Pi^R_{\q}(\omega)}{\partial\omega}
\frac{\partial\Delta^+_{\q}(t,\omega)}{\partial t}+\ldots\ ,\label{imeq}
\end{align}
where the dots indicate neglected higher-order terms.

Consider now an expansion around the equilibrium solution,
\begin{equation}
\Delta^{+}_{\q}(t,\omega) = \Delta^{+}_{\q}(\omega) + 
\delta\Delta^{+}_{\q}(t,\omega)\ .
\end{equation} 
From equation (\ref{imeq}) one reads off
\begin{equation} 
\im\Pi^R_{\q}(\omega)\Delta^{+}_{\q}(\omega)
+\Pi^{+}_{\q}(\omega)\re\Delta^{A}_{\q}(\omega) = 0\ ,
\end{equation}
which is satisfied because of (\ref{ipa}), (\ref{ida}) and the KMS conditions
(\ref{DplusKMS}) and (\ref{PIplusKMS}). 

The first equation (\ref{reeq}) yields for the equilibrium solution, 
\begin{equation}\label{reequil}
\left(\omega^2  -\omega_{\q}^2 -\text{Re}\Pi^R_{\q}(\omega)\right)
\Delta^{+}_{\q}(\omega) = 
- \Pi^{+}_{\q}(\omega)\im\Delta^{A}_{\q}(\omega)\ .
\end{equation}
In the zero-width limit, this equation is fulfilled for
\begin{equation}
\omega = \Omega_{\q} =
\sqrt{\omega_{\q}^2 + \re\Pi^R_{\q}(\Omega_{\q})} \ .
\end{equation}
The finite width leads to a correction,
\begin{equation}
\omega=\Omega_{\q} + \delta\Omega_{\q}\ .
\end{equation}
Expanding (\ref{reequil}) in $\delta\Omega_{\q}$, one obtains to leading
order
\begin{equation}
 2\Omega_{\q}\delta\Omega_{\q}\Delta^+_{\q}(\Omega_{\q})+
\Pi^+(\Omega_{\q})\im\Delta^A_{\q}(\Omega_{\q}) = 0\ ,
\end{equation}
which implies
\begin{equation}\label{do1}
\delta\Omega_{\q} = -\frac{\Gamma_{\q}(\Omega_{\q})}{2}
\frac{\im\Delta^A_{\q}(\Omega_{\q})}{\re\Delta^A_{\q}(\Omega_{\q})}\ .
\end{equation}
We can use the free spectral function,
\begin{equation}
\Delta^-_{\q}(\omega)=2\pi\text{sign}(\omega)\delta(\omega^2-\Omega_{\q}^2)\ ,
\end{equation}
to evaluate $\im\Delta^A_{\q}(\Omega_{\q})$ to leading order in $\Gamma_{\q}$,
\begin{equation}
 \im\Delta^A(\Omega_{\q}) =-
\frac{1}{2\pi}\mathcal{P}\int\frac{\rho(\omega')}{\omega'-\Omega_q}d\omega'
=\frac{1}{4\Omega_{\q}^2}\ .
\end{equation}
Using (\ref{do1}), (\ref{rda}), (\ref{spectralfunction}) and (\ref{widthGamma})
we finally obtain 
\begin{equation}
\delta\Omega_{\q} = \frac{1}{8}\frac{\Gamma_{\q}^2}{\Omega_{\q}}\ . 
\end{equation}
Hence, for $\Gamma_{\q} \ll \Omega_{\q}$, the leading term in the derivative
expansion indeed implies $\omega = \Omega_{\q}$. If finite width effects are 
not negligible, however, off-shell effects become important and the derivative 
expansion becomes unreliable.

Inserting $\omega=\Omega_{\q}$ in the first-order differential equation
(\ref{imeq}), one obtains for the departure from equilibrium of the statistical
propagator,
\begin{equation}\label{deltaDelta}
\left(\left(1 - \frac{1}{2\Omega_{\q}}
\frac{\partial}{\partial \omega}\text{Re}
\Pi^R_{\q}(\omega)\big|_{\Omega_{\q}}\right)\frac{\partial}{\partial t} - 
\frac{1}{\Omega_{\q}}\text{Im}\Pi^R_{\q}(\Omega_{\q})\right)
\delta\Delta^+_{\q}(t;\Omega_{\q}) = 0\ . 
\end{equation}
Hence, $\partial_t\delta\Delta^+_{\q}(t;\Omega_{\q}) = 
{\cal O}(\text{Im}\Pi^R_{\q})$, and to this order Eq.~(\ref{reeq}) is also
satisfied.

We can now evaluate the energy density
\begin{align}
\hat{\epsilon}_{\q}(t) &= \frac{1}{2\omega_{\q}}
\left(\partial_{t_1}\partial_{t_2} 
+ \omega_{\q}^2 \right)\Delta^+_{\q}(t_1,t_2)\big|_{t_1=t_2=t} \nonumber\\
&= \frac{1}{2\omega_{\q}}\int_{-\infty}^{\infty}d\omega
\left(\frac{1}{4}\partial_t^2 + \omega^2 + \omega_{\q}^2\right)
\Delta^{+}_{\q}(t;\omega)\ ,
\end{align}
which approaches the equilibrium value
\begin{equation}
\hat{\epsilon}_{\q}(\infty) \equiv 
\hat{\epsilon}_{\q}^{\text{full}} = 
\frac{\Omega_{\q}^2 + \omega_{\q}^2}{2\omega_{\q}\Omega_{\q}}
\left(\frac{1}{2} + n_\text{B}(\Omega_{\q})\right)\ .
\end{equation}
As already discussed in the previous section, the true equilibrium value of the
energy density does not correspond to a gas of quasi-particles,
\begin{equation}
\hat{\epsilon}_{\q}^{\text{full}} \neq \hat{\epsilon}_{\q}^{\text{QP}}\ .
\end{equation}
From Eqs.~(\ref{deltaDelta}), (\ref{widthGamma}) and (\ref{zGamma}) one 
obtains for the deviation from the equilibrium value,
\begin{equation}
\left(\partial_t + \Gamma_{\q}\right) 
\hat{\epsilon}_{\q}(t) = 0\ ,
\end{equation}
which is identical to the Boltzmann equation (\ref{qpboltz}) for 
quasi-particles. 

In summary, we have obtained the following conditions under which the
approach to equilibrium can be described by Boltzmann equations. For
a dilute, weakly coupled gas the ordinary Boltzmann equation for a
number density is sufficient, which approaches the Bose-Einstein distribution
for a gas of free particles. When interactions with a thermal bath significantly change the spectral function, a Boltzmann equation for quasi-particles
describes the approach to equilibrium as long as the quasi-particle width
can be neglected. However, the equilibrium value of the energy density is
different from the one for a gas of quasi-particles. Finally, when the width
cannot be neglected and off-shell effects become significant, a linear 
evolution equation of `Boltzmann type', which is local in time, is no longer
adequate. Instead, the dynamics is non-local in time, and one has to solve 
Kadanoff-Baym equations.\\

\section{A thermal bath of scalars}\label{ModelSection}

\begin{figure}
  \centering
  \psfrag{1.5}{$1.5$}
\psfrag{1}{$1$}
\psfrag{0.5}{$0.5$}
\psfrag{0}{$0$}
\psfrag{0.8}{$0.8$}
\psfrag{0.6}{$0.6$}
\psfrag{0.4}{$0.4$}
\psfrag{0.2}{$0.2$}
\psfrag{t1}{$T_1$}
\psfrag{t2}{$T_2$}
\psfrag{t3}{$T_3$}
\psfrag{d}{$\rho_{\q}(\omega)$}
\psfrag{omega}{$\omega/m$}
   \includegraphics[width=10cm,height=6cm]{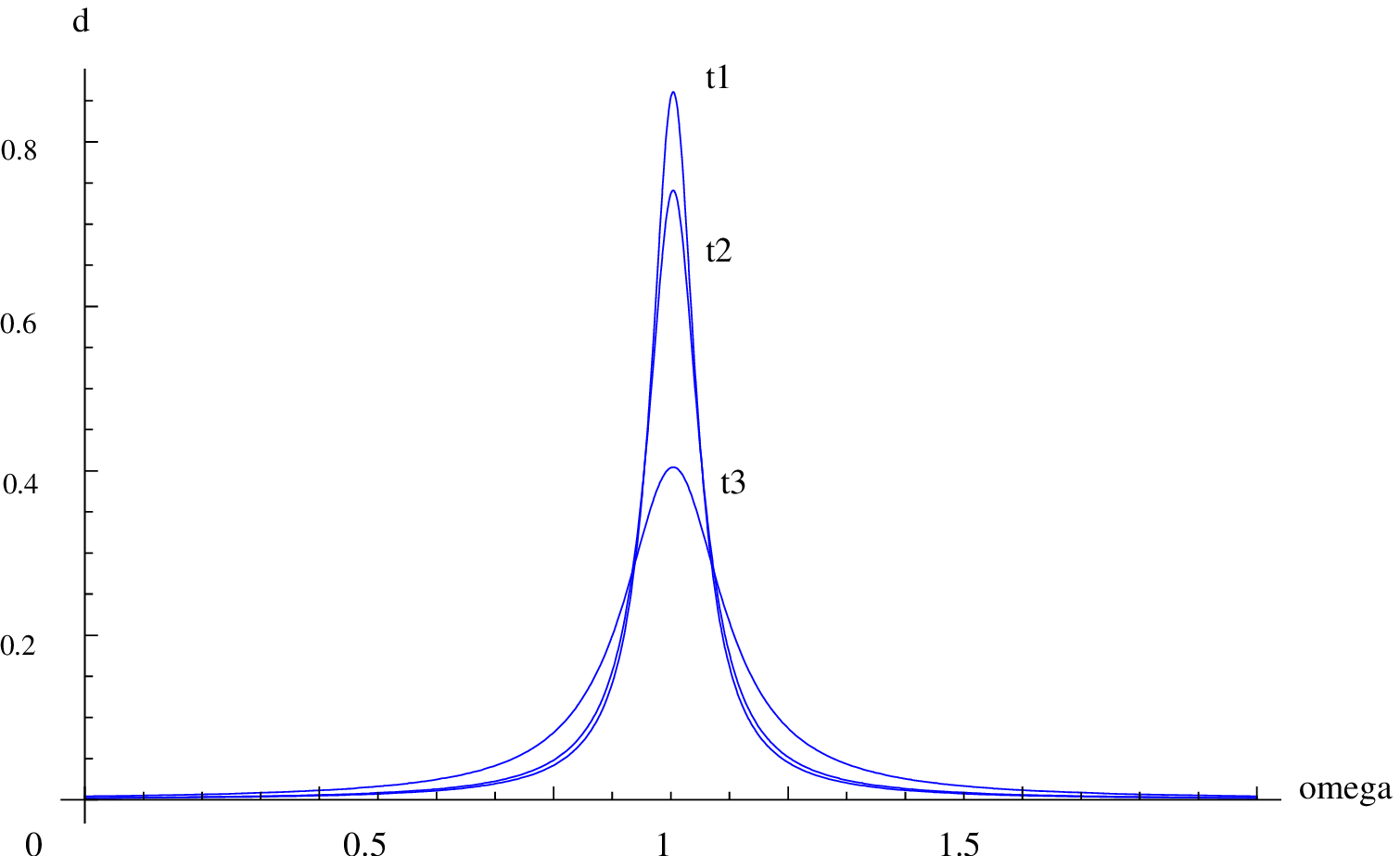}
\caption{Spectral function $\rho_{\q}(\omega)$ for $\textbf{q} =0$; case (a) 
with masses $m_1 = m_2 = 0.2 m$ and temperatures $T_1=0.1 m$, $T_2=0.2 m$, 
$T_3=0.5 m$.} 
\label{spectrala}
\end{figure}
\begin{figure}
  \centering
  \psfrag{t1}{$T_1$}
\psfrag{t2}{$T_2$}
\psfrag{t3}{$T_3$}
\psfrag{800}{$800$}
\psfrag{600}{$600$}
\psfrag{400}{$400$}
\psfrag{200}{$200$}
\psfrag{1.00}{$1$}
\psfrag{0.95}{$0.95$}
\psfrag{0.90}{$0.90$}
\psfrag{0.85}{$0.85$}
\psfrag{delta}{$\rho_{\q}(\omega)$}
\psfrag{omega}{$\omega/m$}
\includegraphics[width=10cm,height=6cm]{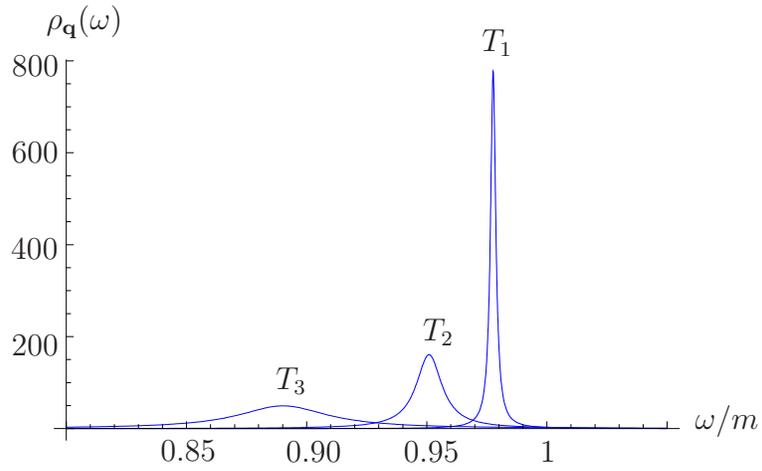}
\caption{Spectral function $\rho_{\q}(\omega)$ for ${\q}=0$; case (b) 
with masses $m_1=m$, $m_2=5m$ and temperatures $T_1=m$, $T_2=2 m$, $T_3=5 m$.}
\label{spectralb}
\end{figure}

So far we have performed a very general analysis, and the only approximation 
has been to neglect the backreaction of the field $\Phi$ on the thermal bath.
Furthermore, we have restricted our discussion to the case that $\Phi$ is 
linearly coupled to the bath via an interaction term $g\Phi\mathcal{O}(\chi)$ 
(cf.~(\ref{action})). In 
general, $\chi$ represents an arbitrary number of bosonic or fermionic fields
with arbitrary couplings including gauge interactions. In order to illustrate
the results of the previous sections, we now consider a toy model 
(cf.~\cite{wel83,bdh04}), where the quanta of two massive scalar fields 
represent the thermal bath. The full Lagrangian is given by
\begin{equation}
\mathcal{L} = 
\frac{1}{2}\partial_{\mu}\Phi\partial^{\mu}\Phi-\frac{1}{2}m^{2}\Phi^{2}
+ \sum_{i=1}^2 \left(\frac{1}{2}\partial_{\mu}\chi_{i}\partial^{\mu}\chi_{i}
-\frac{1}{2}m_{i}^{2}\chi_{i}^{2}\right)
+ g\Phi\chi_{1}\chi_{2}+\mathcal{L}_{\chi \mathrm{int}}\ .
\end{equation}
Note that the coupling $g$ has the dimension of mass. In the following we 
shall neglect self-interaction of the $\chi$ fields and use free thermal 
propagators for simplicity. 

\begin{figure}
  \centering
  \psfrag{20}{$$}
  \psfrag{0}{$0$}
\psfrag{40}{$40$}
\psfrag{60}{$$}
\psfrag{80}{$80$}
  \psfrag{100}{$$}
\psfrag{120}{$120$}
\psfrag{140}{$$}
  \psfrag{delta}{$\Delta^-_{\q}(y)$}
\psfrag{omega}{$my$}
\psfrag{1.0}{$1.0$}
\psfrag{0.5}{$0.5$}
\psfrag{0.6}{$-0.5$}
\psfrag{1.1}{$-1.0$}
   \includegraphics[width=10cm,height=8cm]{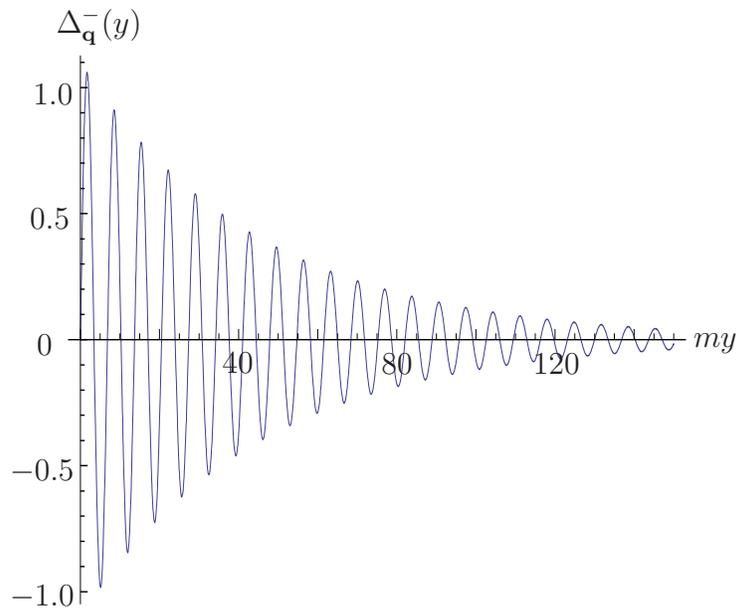}
  \caption{Spectral function $\Delta^-_{\q}(y)$ for ${\q}=0$; case (b) with
masses $m_1=m$, $m_2=5m$ and $T=10m$.}
  \label{Delta-}
\end{figure}

We consider two cases: (a) $m \gg m_1,m_2$ and (b) $m_2 \gg m,m_1$. In the 
first case, dissipation is dominated by $\Phi$ decays and inverse decays, 
$\Phi \leftrightarrow \chi_1\chi_2$, whereas in the second one $\chi_2$ decays 
and inverse decays, $\chi_{2}\leftrightarrow\Phi\chi_{1}$, are most important.
\begin{figure}
  \centering
\psfrag{RePi}{$\re\Pi^R_{\q}(\omega)$}
\psfrag{1.0}{$1.0$}
\psfrag{0.5}{$0.5$}
\psfrag{0}{$0$}
\psfrag{1.5}{$1.5$}
\psfrag{2.0}{$2.0$}
\psfrag{0.2}{$-0.2$}
\psfrag{0.4}{$-0.4$}
\psfrag{0.6}{$-0.6$}
\psfrag{0.8}{$-0.8$}
\psfrag{omega}{$\omega/m$}
   \includegraphics[width=10cm,height=8cm]{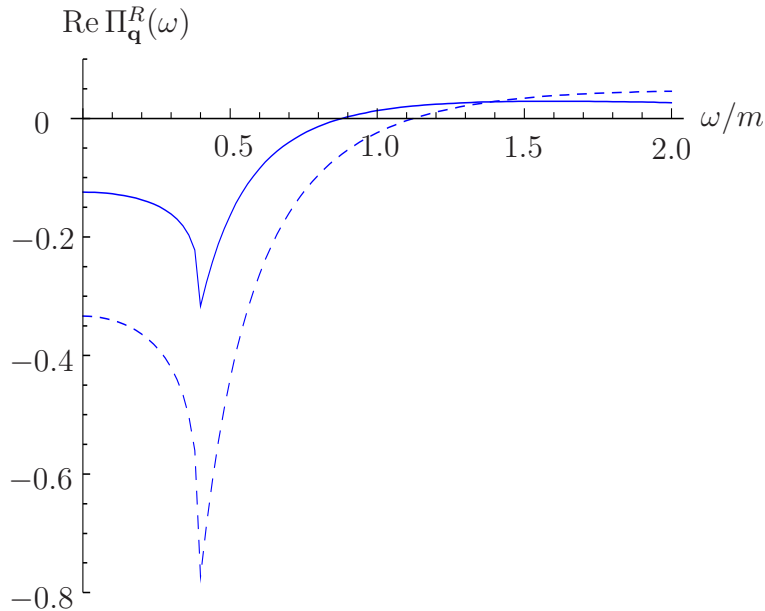}
  \caption{Real part of the self-energy $\Pi^R_{\q}(\omega)$ for ${\q}=0$; case (a) with masses $m_1=m_2=0.2m$ and temperatures $T_1 = 0.5m$ (solid) 
and $T_2 = m$ (dashed).}
  \label{repi}
\end{figure}

In both cases the imaginary part of the self-energy is known analytically 
\cite{bdh04,dre06}. The relevant formulae are collected in 
Appendix \ref{AnalyticStructure}. For $m \gg m_1,m_2$, the decay width
of $\Phi$ at zero temperature is given by
\begin{equation}
\Gamma = \frac{1}{16\pi}\left(\frac{g}{m}\right)^2 m \ .
\end{equation}
To illustrate thermal effects we shall use a rather large coupling which
corresponds to $\Gamma/m = 0.1$.

The spectral function $\rho_{\q}(\omega)$ (cf.~(\ref{spectralfunctionReno}))
is shown in Figures~\ref{spectrala} and \ref{spectralb}  for the two mass 
patterns (a) and (b), respectively. 
In case (a), $\Pi^R_{\q}$ has an imaginary part at zero temperature. The
width is large, and already at small temperatures the quasi-particle profile
becomes broad. On the contrary, in case (b) the zero-temperature width is
zero and the finite-temperature width is small. Hence, the quasi-particle
profile becomes broad only at much larger temperatures. The spectral function
$\Delta^-_{\q}(y)$ is the Fourier transform of $i\rho_{\q}(\omega)$. As Figure
\ref{Delta-} illustrates, it approximately represents a damped 
oscillation with frequency $\Omega_{\q}$ and damping rate $\Gamma_{\q}$.

It is interesting that thermal corrections can increase or decrease the
particle mass $m$. Whether the quasi-particle peak moves to the right or to 
the left depends on the position of the zero-temperature pole relative to the 
branch cuts, and it also depends on the temperature. This can be seen by 
considering the real part of the self-energy, which is displayed for two 
different temperatures in Fig.~\ref{repi} for case (a). For the smaller
temperature one has $\re\Pi^R_{\q=0}(m) > 0$, whereas $\re\Pi^R_{\q=0}(m) < 0$ 
holds for the larger temperature, which corresponds to a shift of the particle
mass to the right and to the left, respectively. 

The statistical propagator $\Delta^{+}_{\q}(t_1,t_2)$ depends on the initial
conditions. The most general gaussian initial density matrix has five free 
parameters (cf.~\cite{ber04}). We consider the simplest case of a free field 
density matrix and vanishing mean values $\Phi$ and $\dot{\Phi}$, which implies 
for each momentum mode,
\begin{align}
\Phi_{\q,\text{in}}&=0 \ ,\\
\dot{\Phi}_{\q,\text{in}}&=0 \ ,\\
\Delta^{+}_{\q,\text{in}} &= 
\Delta^{+}_{\q}(t_{1},t_{2})|_{t_{1}=t_{2}=0} = \frac{1}{\omega_{\q}}
\left(\frac{1}{2}+n_{\q}\right)\ ,\label{in1}\\
\dot{\Delta}^{+}_{\q,\text{in}} &= 
\partial_{t_{1}}\Delta^{+}_{\q}(t_{1},t_{2})|_{t_{1}=t_{2}=0} 
= \partial_{t_{2}}\Delta^{+}_{\q}(t_{1},t_{2})|_{t_{1}=t_{2}=0} = 0\ , 
\label{in2}\\
\ddot{\Delta}^{+}_{\q,\text{in}} &= \partial_{t_{1}}\partial_{t_{2}} 
\Delta^{+}_{\q}(t_{1},t_{2})|_{t_{1}=t_{2}=0} = \omega_{\q}
\left(\frac{1}{2}+n_{\q}\right)\ .\label{in3}
\end{align}
The initial state of the system is now characterized by only one parameter
$n_{\q}$ which corresponds to an initial number density for a free field.

\begin{figure}[t]
  \centering
\psfrag{t1}{$mt_1$}
\psfrag{t2}{$mt_2$}
 \psfrag{40}{$20$}
\psfrag{20}{$10$}
\psfrag{60}{$30$}
\psfrag{10}{$10$}
\psfrag{-10}{$10$}
\psfrag{0}{$0$}
\psfrag{delta}{$\Delta^+_{\q}$}
\includegraphics[width=12cm]{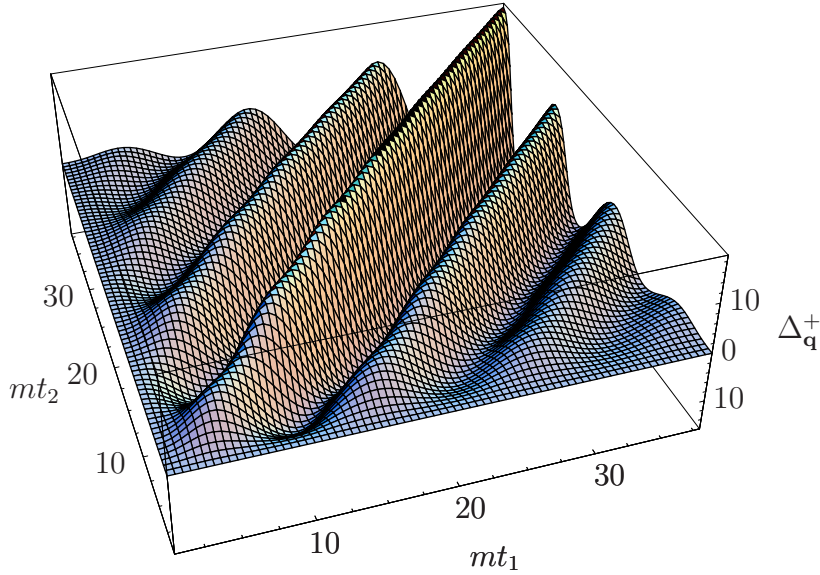}
\caption{Statistical propagator $\Delta^+_{\q}(t_1,t_2)$ for ${\q}=0$; case (b) with
masses $m_1=m$, $m_2=5m$ and
$T=10m$.} 
\label{delta+full}
\end{figure}
\begin{figure}
  \centering
 \psfrag{15}{$15$}
\psfrag{10}{$10$}
\psfrag{5}{$5$}
\psfrag{0}{$0$}
 \psfrag{20}{$20$}
\psfrag{25}{$25$}
\psfrag{30}{$30$}
 \psfrag{51}{$-5$}
\psfrag{11}{$-10$}
\psfrag{16}{$-15$}
\psfrag{delta}{$\Delta_{\q}^+(t_1,t_2)$}
\psfrag{omega}{$my$}
   \includegraphics[width=10cm,height=8cm]{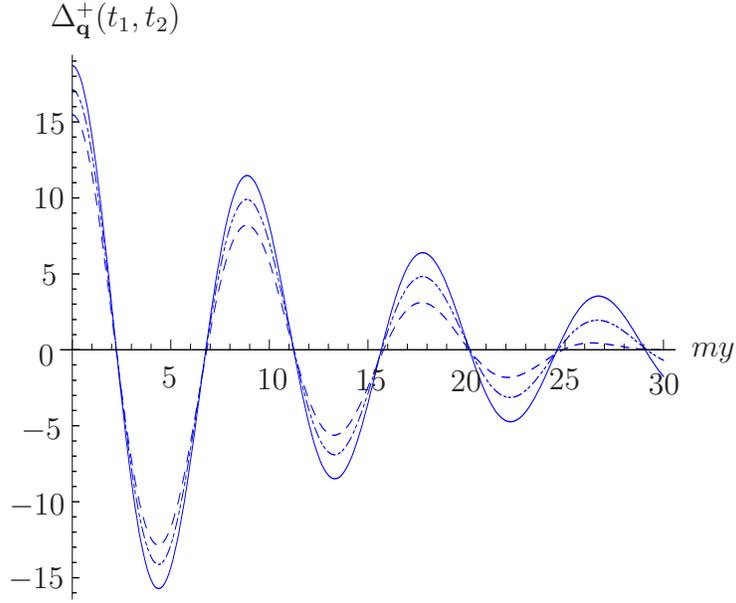}
  \caption{Statistical propagator $\Delta_{\q}^+(t_1,t_2)$ as function of 
$y=t_1-t_2$ for ${\q}=0$; case (b) with $m_1=m$, $m_2=5m$, $T=10m$ and three values of $t=(t_1+t_2)/2$: $mt=15$ (dashed line), $mt=20$ (dotted-dashed), $mt=60$ (solid).}
  \label{delta+y}
\end{figure}
\begin{figure}
  \centering
   \psfrag{5}{$5$}
\psfrag{10}{$10$}
\psfrag{5}{$5$}
\psfrag{0}{$0$}
 \psfrag{15}{$15$}
\psfrag{20}{$20$}
\psfrag{100}{$100$}
 \psfrag{40}{$40$}
\psfrag{140}{$140$}
\psfrag{60}{$60$}
  \psfrag{delta}{$\Delta_{\q}^+(t_1,t_2)$}
\psfrag{omega}{$mt$}
   \includegraphics[width=10cm,height=7.5cm]{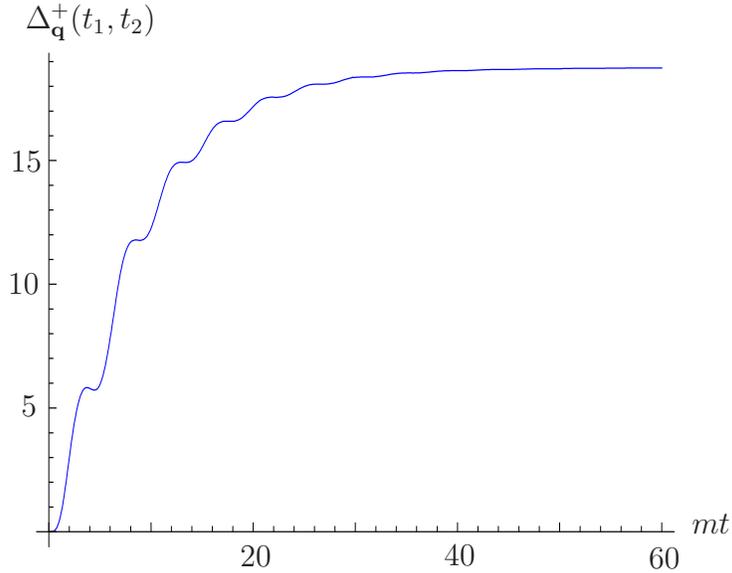}
  \caption{Statistical propagator $\Delta_{\q}^+(t_1,t_2)$ as function of 
$t=(t_1+t_2)/2$ for $y=0$, ${\q}=0$; case (b) with masses $m_1=m$, $m_2=5m$ and $T=10m$.}
  \label{delta+t}
\end{figure}
\begin{figure}
  \centering
   \psfrag{5}{$5$}
\psfrag{10}{$10$}
\psfrag{5}{$5$}
 \psfrag{15}{$15$}
\psfrag{20}{$20$}
\psfrag{0}{$0$}
\psfrag{25}{$25$}
\psfrag{30}{$30$}
\psfrag{35}{$35$}
 \psfrag{40}{$40$}
\psfrag{50}{$50$}
\psfrag{60}{$60$}
  \psfrag{delta}{$\Delta_{\q}^+(t_1,t_2)$}
\psfrag{tm}{$mt$}
   \includegraphics[width=10cm,height=8cm]{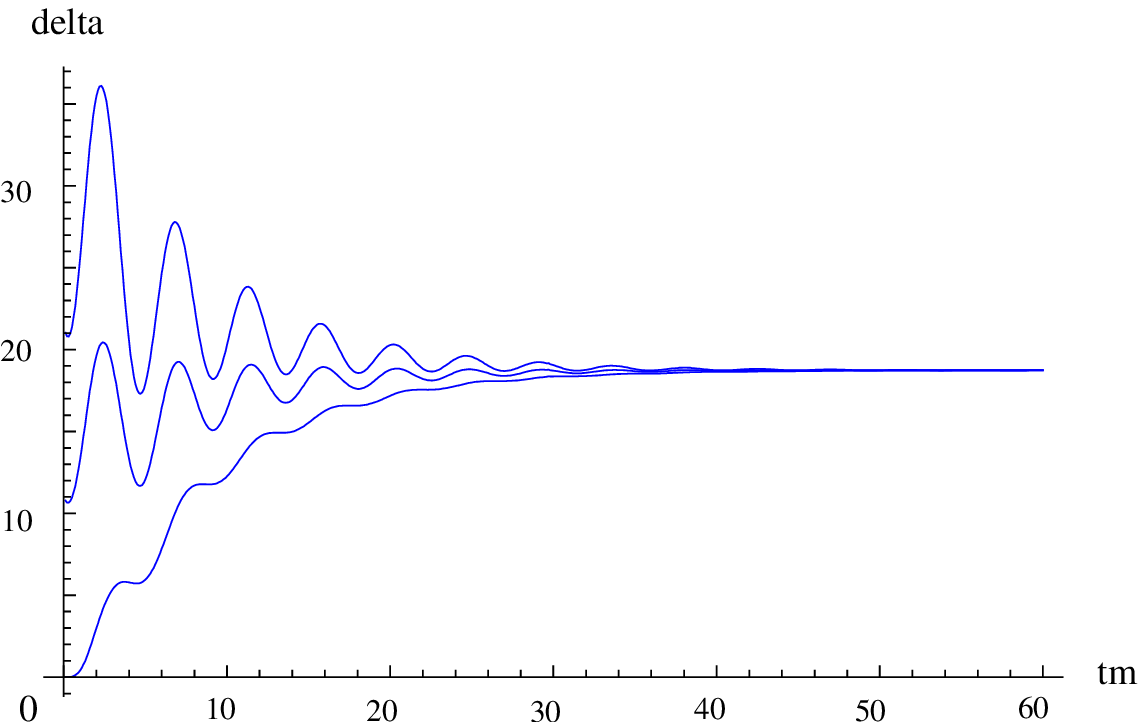}
\caption{Statistical propagator $\Delta_{\q}^+(t_1,t_2)$ with $\textbf{q}=0$ as function of 
$t=(t_1+t_2)/2$ for $y=0$ and different initial conditions; case (b) with masses 
${\q}=0$, $m_1=m$, $m_2=5m$ and $T=10m$.}
  \label{delta+ini}
\end{figure}

The general solution (\ref{solution}) for the statistical propagator
$\Delta^+_{\q}(t_1,t_2)$ is shown in Figure~\ref{delta+full}. For fixed 
$t=(t_1+t_2)/2$ one sees damped oscillations in $y=t_1-t_2$. The amplitude 
increases with increasing time $t$, as illustrated by Figure~\ref{delta+y}. 
For fixed $y=t_1-t_2$ one observes the approach to equilibrium with increasing
$t=(t_1+t_2)/2$. For large times the departure from equilibrium is described
by a first-order differential equation, and it decreases exponentially.
At small times the evolution is governed by a second-order differential
equation, which leads to the oscillations visible in Figure~\ref{delta+t}.
The independence of the equilibrium solution from the initial conditions
is illustrated by Figure~\ref{delta+ini}. The memory of the initial conditions
is lost at times $t>1/\Gamma$.

\begin{figure}
  \centering
\psfrag{e}{$\begin{Large}\hat{\epsilon}_{\q}\end{Large}$}
\psfrag{2}{$2$}
\psfrag{4}{$4$}
\psfrag{6}{$6$}
\psfrag{8}{$8$}
\psfrag{10}{$10$}
\psfrag{0}{$0$}
\psfrag{T}{$T/m$}
\includegraphics[width=10 cm]{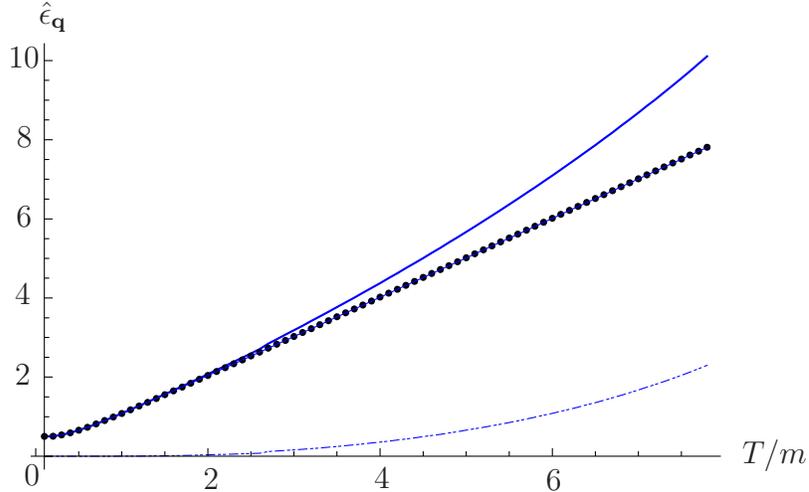}
\caption{Energy density $\hat{\epsilon}_{\q}=\epsilon_{\q}/\omega_{\q}$ as 
function of temperature for ${\q}=0$; case (b) with
masses $m_1=m$, $m_2=5m$: total energy density (solid), particle 
and quasi-particle energy densities (dotted), and 'vacuum' energy density (dashed).} 
\label{equilibriumenergy}
\end{figure}

Finally, it is important to recall that the equilibrium value of the energy
differs from the one obtained in the Boltzmann approximation. This is 
illustrated in Figure~\ref{equilibriumenergy} where the different contributions
to the energy are compared as functions of temperature. For the chosen
parameters the particle and quasi-particle energies are indistinguishable.
The `vacuum contribution' is positive, which means that the total energy
is larger than the particle/quasi-particle one. The reason is that for the
chosen parameters thermal corrections decrease the particle mass. For other
parameter choices the `vacuum contribution' can have opposite sign.

\section{Conclusions and outlook}

We have studied the approach to equilibrium for a real scalar field coupled
to a large thermal bath. We have computed the exact two-point functions, the 
spectral function and the statistical propagator, for arbitrary initial 
conditions. This is possible for a thermal bath with many degrees of freedom 
such that the backreaction of the scalar field can be neglected.

The self-energy representing the thermal bath is time-translation invariant.
We have shown that this is also the case for the spectral function, whereas
the statistical propagator depends on two time coordinates, $t_1$ and $t_2$, 
and also the time $t_{i}$ where the initial conditions are specified.

We have obtained the two-point functions by solving the Kadanoff-Baym 
equations, which turned out to be equivalent to solving a stochastic
Langevin equation. As expected, the relaxation time is determined by the
imaginary part of the self-energy, i.e., a `quasi-particle width' $\Gamma$. 
For $t > 1/\Gamma$, the statistical propagator becomes independent of the
initial conditions. It is then given by a memory integral which depends 
on the real and imaginary part of the self-energy. 

As long as thermal corrections are small, the approach to equilibrium is
well described by the ordinary Boltzmann equation, which is a local, 
first-order differential equation in time for the particle number density. 
However,
in the case of large thermal corrections the notion of number density becomes ambiguous, and it is important to consider the second-order Kadanoff-Baym
equations for the two-point functions rather than a Boltzmann equation. 
Still, as long as the quasi-particle decay width is small compared to the
quasi-particle energy, a Boltzmann equation for quasi-particles describes
the approach to equilibrium to good approximation. For large decay width
the dynamics becomes nonlocal in time and the Boltzmann approximation breaks
down.
 
It is interesting to study the contribution of the thermalized scalar field
to energy density and pressure. For a free field these observables are 
determined by the Bose-Einstein distribution function. Interaction with the 
thermal bath can significantly modify energy density and pressure, and 
therefore the equation of state. The Bose-Einstein distribution as function 
of the complex quasi-particle pole is now the relevant quantity. Energy 
density and pressure differ from the
expressions for a free gas of quasi-particles by a temperature-dependent
`vacuum term' which can become important at high temperatures. 

We have illustrated these results for a toy model where the thermal bath
consists of two massive scalar fields. We have considered two cases where
equilibration takes place either via decays and inverse decays or via
`Landau damping'. In general, one has to study the Kadanoff-Baym equation
for the statistical propagator which depends on two time coordinates as well 
as initial conditions. However, for large times, $t \gg 1/\Gamma$, the 
time evolution is well described by the Boltzmann equation for quasi-particles.

Our analysis has been motivated by the need of a full quantum mechanical
description of leptogenesis. To achieve this, one has to consider correlation 
functions rather than number densities, although for parts of the calculation 
the use of Boltzmann equations will be sufficient. 
The heavy Majorana neutrino is very weakly coupled to the thermal bath. 
Hence, thermal corrections to its mass and width are small, and its
approach to equilibrium is well described by Boltzmann equations. However,
to study the dependence of the final baryon asymmetry on initial conditions
it may be necessary to consider the statistical propagator, since 
leptogenesis takes place at $t_B \sim 1/\Gamma$. Furthermore, for lepton and 
Higgs fields, which have strong gauge interactions, finite-width effects can 
be important. At present it is unclear how accurately the leptogenesis
process can be described based on a quasi-particle picture for the standard
model particles which form the thermal bath. These questions are currently 
under investigation \cite{abx09}.\\

\vspace{1cm}
\noindent
{\bf\large Acknowledgements}\\

\noindent
We would like to thank J.~Berges, D.~B\"odeker, J.~Schmidt and C.~Wetterich
for helpful discussions.

\newpage
\begin{appendix}

\section{The spectral function}\label{Aspectralfunction}

\subsection{Time-translation invariance}\label{TimeTranslInv}

In this section we shall prove that the most general solution of the first
Kadanoff-Baym equation is time-translation invariant. The starting point is
Eq.~(\ref{kbe1}) with the boundary conditions (\ref{cond1}) - (\ref{cond3}).
Performing the change of variables $t_1=t+y/2$, $t_2=t-y/2$, Eq.~(\ref{kbe1})
becomes
\be\label{kbe1ty}
\left(\frac{1}{4}\partial^2_t+\partial_t\partial_y+\partial^2_y
+\omega_{\q}^2\right)\Delta_{\q}^-(t;y)+\int_{0}^{y}dy'\Pi_{\q}^-(y-y')
\Delta_{\q}^-(t';y') = 0\ ,
\ee
where $t'=t-(y-y')/2$ and $\omega_{\q}^2={\q}^2+m^2$. Note that $\Delta_{\q}^-$
and $\Pi^-_{\q}$ only depend on $|{\q}|$ because of rotational invariance. 
Both functions are antisymmetric in $y$. The boundary conditions 
(\ref{cond1}) - (\ref{cond3}) read
\begin{align}
\Delta^{-}_{\q}(t;0)&= 0\ , \label{co1}\\
\partial_t\Delta^{-}_{\q}(t;0)&= 0\ ,\label{co2}\\
\partial_y\Delta^{-}_{\q}(t;y)|_{y=0} &= 1\ , \label{co3}\\
\left(\frac{1}{4}\partial_t^2 - \partial_y^2\right)
\Delta^{-}_{\q}(t;y)|_{y=0} &= 0\ .
\label{co4}
\end{align}
The condition ($\ref{co1}$) is automatically fulfilled because of the 
antisymmetry in $y$. 

To prove that $\Delta^-$ is time-translation invariant we now perform an
expansion in powers of $\Pi^-$,
\be
\Delta^-_{\q}=\sum_{n=0}^{\infty}\Delta^{(n)}_{\q}\ , \quad
\Delta^{(n)}_{\q}=\mathcal{O}(\Pi_{\q}^{(n)})\ .
\ee
For $n=0$ one has
\be\label{kadaf1}
\left(\frac{1}{4}\partial^2_t + \partial_t\partial_y + \partial^2_y
+\omega_{\q}^2\right)\Delta_{\q}^{(0)}(t;y) = 0\ .
\ee
Using the antisymmetry of $\Delta^-_{\q}$ in $y$, one obtains 
\be\label{kadaf2}
\partial_t\partial_y\Delta^{(0)}_{\q}(t;y) = 0\ ,
\ee
which has the general solution 
\be
\Delta_{\q}^{(0)}=a^{(0)}_{\q}(t)+b^{(0)}_{\q}(y)\ .
\ee
Every solution of Eqs.~(\ref{kadaf1}) and (\ref{kadaf2}) satisfies the
boundary condition (\ref{co4}). The condition (\ref{co2}) implies
\be
\partial_t\Delta_{\q}^{(0)}(t;0) = \partial_t a^{(0)}_{\q}(t) = 0 \ .
\ee
Hence, $a^{(0)}_{\q}$ is constant and $\Delta_{\q}^{(0)}$ only depends on $y$.
Eq.~(\ref{kadaf1}) now becomes
\be
\left(\partial^2_y + \omega_{\q}^2\right)\Delta_{\q}^{(0)}(y) = 0\ ,
\ee
which has the antisymmetric solution
\be
\Delta^{(0)}_{\q} (y) = c^{(0)}_{\q}\sin(\omega_{\q}y)\ . 
\ee
For $n\ne0$ one can use the recurrence relation
\be\label{bay}
\left(\frac{1}{4}\partial^2_t + \partial_t\partial_y + \partial^2_y
+ \omega_{\q}^2\right)\Delta_{\q}^{(n+1)}(t;y) +
\int_{0}^{y}dy'\Pi_{\q}^-(y-y')\Delta_{\q}^{(n)}(y') = 0\ .
\ee
Using the antisymmetry of $\Pi^-_{\q}$ and $\Delta^-_{\q}$ in $y$, one again
finds
\be
\partial_t\partial_y\Delta^{(n+1)}_{\q}(t;y) = 0 \ .
\ee
Repeating the same steps as for $\Delta^{(0)}_{\q}$ yields the result that also
$\Delta^{(n+1)}_{\q}$ is independent of $t$.

We conclude that the spectral function is the antisymmetric solution of
the equation 
\be\label{kbe1f}
\left(\partial^2_y + \omega_{\q}^2\right)\Delta_{\q}^-(y) + 
\int_{0}^{y}dy'\Pi_{\q}^-(y-y')\Delta_{\q}^-(y') = 0\ ,
\ee
with the boundary condition
\be
\partial_y\Delta^{-}_{\q}(y)|_{y=0} = 1\ .
\ee

\subsection{Conventions for propagators and self-energies}

In thermal equilibrium the retarded and advanced propagators and self-energies
only depend on the time difference $y=t_1-t_2$. In the following we list
several relations between their Fourier transforms, which are used in the
different sections. In principle, these relations are all well know, but
their specific form depends on the chosen conventions. All relations are
not affected by the three-dimensional Fourier transform. We therefore drop
the argument ${\q}$ or ${\bf x}$. \\

\noindent
Propagators:
\begin{eqnarray}
\Delta^-(\omega)^* &=& -\Delta^-(\omega)\ , \label{dm} \\
\Delta^+(\omega)^* &=& \Delta^+(\omega)\ , \label{dp} \\
\Delta^A(\omega) &=& \frac{i}{2}\Delta^-(\omega)
- \mathcal{P}\int_{-\infty}^{\infty}\frac{d\omega'}{2\pi}
\frac{\Delta^-(\omega')}{\omega'-\omega}\ , \label{da} \\
\Delta^R(\omega) &=& -\frac{i}{2}\Delta^-(\omega)
- \mathcal{P}\int_{-\infty}^{\infty}\frac{d\omega'}{2\pi}
\frac{\Delta^-(\omega')}{\omega'-\omega}\ , \label{dr} \\
\re\Delta^A(\omega) &=& -\re\Delta^R(\omega)=\frac{i}{2}\Delta^-(\omega)\ ,
\label{rda}\\
\im\Delta^A(\omega) &=& \im\Delta^R(\omega)
= -\mathcal{P}\int_{-\infty}^{\infty}\frac{d\omega'}{2\pi i}
\frac{\Delta^-(\omega')}{\omega'-\omega}\ , \label{ida}\\
\Delta^A(-\omega) &=& \Delta^R(\omega)\ .\label{dar}\\
\end{eqnarray}

\noindent
Self-energies:
\begin{eqnarray}
\Pi^-(\omega)^* &=& -\Pi^-(\omega)\ ,\label{pm}\\
\Pi^+(\omega)^* &=& \Pi^+(\omega)\ ,\label{pp}\\
 \Pi^A(\omega) &=& -\frac{1}{2}\Pi^-(\omega) +
\mathcal{P}\int\frac{d\omega'}{2\pi i}
\frac{\Pi^-(\omega')}{\omega'-\omega}\ ,\label{pa}\\
\Pi^R(\omega) &=& \frac{1}{2}\Pi^-(\omega) + 
\mathcal{P}\int\frac{d\omega'}{2\pi i}
\frac{\Pi^-(\omega')}{\omega'-\omega}\ ,\label{pr}\\
\re\Pi^A(\omega) &=& \re\Pi^R(\omega) = \mathcal{P}\int\frac{d\omega'}{2\pi i}
\frac{\Pi^-(\omega')}{\omega'-\omega}\ ,\label{rpa}\\
\im\Pi^A(\omega) &=& -\im\Pi^R(\omega) = \frac{i}{2}\Pi^-(\omega)\ ,
\label{ipa}\\
\Pi^A(-\omega) &=& \Pi^R(\omega)\label{par}\ .
\end{eqnarray}

\subsection{Scalar field model}\label{AnalyticStructure}

The interaction with the thermal bath changes the spectral function of a free
scalar particle,
\begin{equation}\label{freespec}
\rho_{\q}(\omega)=2\pi \text{sign}(\omega)\delta(\omega^{2}-\omega_{\q}^{2})\ ,
\end{equation}
to the expression (\ref{spectralfunctionReno}) which depends on real and 
imaginary part of the self-energy, 
\be
\rho_{\q}(\omega)={-2{\rm Im}\Pi^R_{\q}(\omega)+2\omega\epsilon\over 
[\omega^2-\omega_{\q}^2-{\rm Re}\Pi^R_{\q}(\omega)]^2+
[{\rm Im}\Pi^R_{\q}(\omega)+\omega\epsilon]^2}\ .\label{Aspectralfunction}
\ee

We have computed the imaginary part of the self-energy in the scalar field 
model defined in Section~\ref{ModelSection}, assuming free thermal
propagators for the fields $\chi_1$ and $\chi_2$. The result agrees
with \cite{bdh04}. One obtains ($q=(\omega_{\q},\q)$):
\begin{equation}
-\im\Pi^{R}_{\q}(\omega) = \sigma_0(q)+\sigma^{(a)}_{\beta}(q)
+\sigma^{(b)}_{\beta}(q)\ .
\end{equation}
Here $\sigma_{0}$ is the zero-temperature contribution due to the decay 
process $\Phi\rightarrow\chi_{1}\chi_{2}$,
\begin{align}
\sigma_{0}(q) = &\frac{g^{2}}{16\pi q^{2}}\text{sign}(\omega)
\Theta(q^{2}-(m_{1}+m_{2})^{2})\nonumber\\
&\times \left((q^{2})^{2}-2q^{2}(m_{1}^{2}+m_{2}^{2})
+(m_{1}^{2}-m_{2}^{2})^{2}\right)^{\frac{1}{2}}\ ,
\end{align}
$\sigma^{(a)}_{\beta}$ is the finite-temperature contribution from this 
process,
\begin{align}
\sigma^{(a)}_{\beta}(q) = &\frac{g^{2}}{16\pi|\q|\beta}
\text{sign}(\omega) \Theta(q^{2}-(m_{1}+m_{2})^{2})\nonumber\\
&\times \left(\ln\left(\frac{1-e^{-\beta\omega_{+}}}{1-e^{-\beta\omega_{-}}}
\right)+ (m_{1}\leftrightarrow m_{2})\right)\ ,
\end{align}
and $\sigma^{(b)}_{\beta}(\q)$ is the finite-temperature contribution from 
processes $\chi_{i}\rightarrow\chi_{j}\phi$,\footnote{Note that we disagree with the discussion 
in \cite{wel83} which implies the additional factor $\Theta(|m_1^2-m_2^2|-q^2)$.}
\begin{align}
\sigma^{(b)}_{\beta}(q)= &\frac{g^{2}}{16\pi|\q|\beta}
\text{sign}(\omega) \Theta((m_{1}-m_{2})^{2}-q^{2})\nonumber\\
&\times\left(\ln\left(\frac{1-e^{-\beta|\omega_{-}|}}{1-e^{-\beta|\omega_{+}|}}
\right)+(m_{1}\leftrightarrow m_{2})\right)\ ,
\end{align}
where we have used the abbreviations
\begin{equation}
\omega_{\pm} = \frac{|\omega|}{2q^{2}}(q^{2}+m_{1}^{2}-m_{2}^{2})
\pm\frac{|\textbf{q}|}{2|q^{2}|}\left( 
(q^{2}+m_{1}^{2}-m_{2}^{2})^{2}-4q^{2}m_{1}^{2}\right)^{\frac{1}{2}}\ .
\end{equation}
The real part of the self-energy can be computed using the dispersion relation 
relation,
\begin{equation}
\re\Pi^R_{\q}(\omega)=\frac{1}{\pi}\mathcal{P}\int_{-\infty}^{\infty}d\omega'
\frac{\im\Pi^R_{\q}(\omega')}{\omega'-\omega}\ .
\end{equation}

\begin{figure}[htbp]
  \centering
   \includegraphics[width=12 cm]{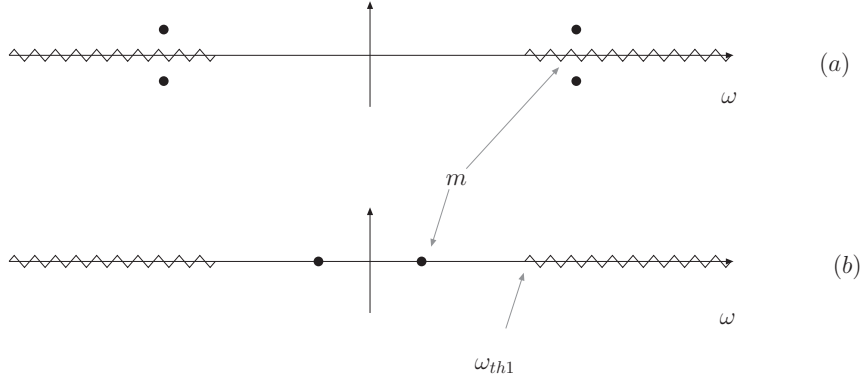}
\caption{Poles and cuts of the spectral function $\rho(\omega)$ for $\textbf{q}=0$ at $T=0$:
(a) $m > m_1+m_2$, and (b) $m < m_1 + m_2$.}\label{spectralstructureT0}
\end{figure}

Based on these expressions we can discuss the analytic structure of the 
spectral function. For a free field $\rho_{\q}(\omega)$ is given by 
(\ref{freespec}) which has two poles at $\omega = \pm \omega_q$ in the complex
$\omega$-plane. The interaction of $\Phi$ with $\chi_1$ and $\chi_2$ does 
not modify these poles for $m < m_1 + m_2$, where $\Phi$ is stable at zero
temperature. In addition there are branch cuts at the two-particle thresholds
$|\omega| > \omega_{\text{th}1} = \sqrt{{\q}^{2}+(m_{1}+m_{2})^{2}}$ 
(see Fig.~\ref{spectralstructureT0}$b$). They correspond to
virtual decays and inverse decays, $\Phi\leftrightarrow\chi\chi$. 
In the case $m>m_{1}+m_{2}$ these 
processes can happen on-shell since $m>\omega_{\text{th}1}$, and $\Phi$ becomes 
unstable. Now the spectral function has four poles in the complex 
$\omega$-plane, whose real parts lie in the region of the branch cuts
(see Fig.~\ref{spectralstructureT0}$a$).
The imaginary parts of the poles correspond to the decay width of $\Phi$.

\begin{figure}
  \centering
\includegraphics[width=12 cm]{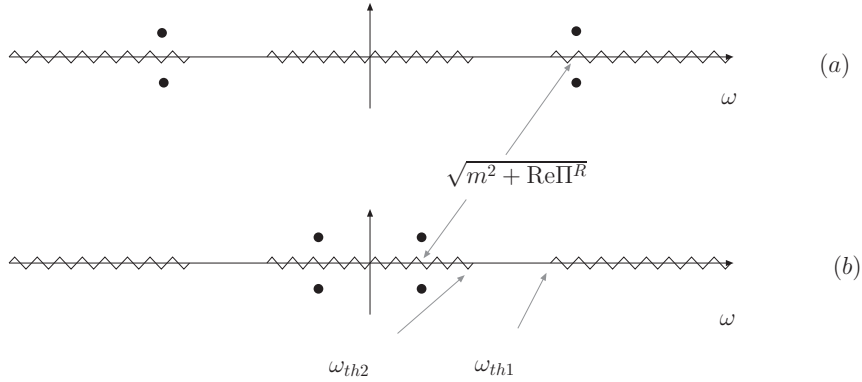}
\caption{Poles and cuts of the spectral function $\rho(\omega)$ for $\textbf{q}=0$ at 
$T\neq 0$: (a) $m > m_1+m_2$, and (b) $m < m_1 + m_2$.}
\label{spectralstructureT}
\end{figure}

The analytic structure of the spectral function at finite temperature is
displayed in Fig.~\ref{spectralstructureT}. The position of 
$\omega_{\text{th}1}$ is shifted due to thermal corrections from 
$\re\Pi^{R}_{\q}$. Furthermore, a new branch cut appears in the region where 
$\sigma_{b}\neq 0$, i.e. for 
$|\omega|<\omega_{\text{th}2} = \sqrt{{\q}^2 + (m_{1}-m_{2})^{2}}$. This is 
due to processes $\chi\leftrightarrow\phi\chi$ and corresponds to Landau 
damping of quasi-particles in the plasma. If the real part of the poles
falls into the regions of one of the branch cuts, i.e. 
$|\omega|<\omega_{\text{th}2}$ or $|\omega|>\omega_{\text{th}1}$, they
acquire an imaginary part which corresponds to the quasi-particle decay 
width (see Fig.~\ref{spectralstructureT}). 
Qualitatively, this analytic structure is typical for interacting quantum 
field theories at finite temperature. In general, the spectral function can 
have additional singular contributions for $m <|\omega|<\omega_{\text{th}1}$ 
corresponding to bound states. At finite temperature they are also dressed to 
quasi-particles.

\subsection{Breit-Wigner approximation}\label{ABreitWigner}

In the regime of couplings and temperatures where 
$|\im\Pi^R_{\q}(\Omega_{\q})| \ll \Omega_{\q}^{2}$,
so that the quasi-particle picture holds, one can approximate the spectral
function $\rho_{\q}(\omega)$ by a Breit-Wigner function. From the expression
(\ref{spectralfunctionReno}) one easily obtains
\begin{equation}\label{BWapp}
\rho_{\q}(\omega)\simeq \frac{Z_{\q}}{2\Omega_{\q}}
\frac{\text{sign}(\omega)\Gamma_{\q}}
{\left(|\omega|-\Omega_{\q}\right)^{2}
+\frac{1}{4}\Gamma_{\q}^{2}}\ ,
\end{equation}
where $\Gamma_{\q}$ is the quasi-particle width 
\begin{equation}\label{widthGamma}
\Gamma_{\q} = -Z_{\q}\frac{{\im}\Pi^{R}_{\q}(\Omega_{\q})}{\Omega_{\q}}\ ,
\end{equation}
with
\begin{equation}\label{zGamma}
Z_{\q}=\left(1-\frac{1}{2\Omega_{\q}}
\frac{\partial{\re}\Pi^{R}_{\q}(\omega)}{\partial\omega}
\Big|_{\omega=\Omega_{\q}}\right)^{-1}\ .
\end{equation}

Contrary to the exact spectral function (\ref{spectralfunctionReno}), the
Breit-Wigner approximation (\ref{BWapp}) has no branch cuts. The integrals over $\omega$ are dominated by the regions around the 
quasi-particle poles where the two functions are very similar. For the
Fourier transform, the spectral function in real time, one obtains
\begin{equation}\label{DeltaMinusBW}
\Delta^{-}_{\q}(y)\simeq Z_{\q}\frac{\sin(\Omega_{q}y)}{\Omega_{\q}}
e^{-\Gamma_{q}t/2}\ .
\end{equation}

\end{appendix}


\clearpage


\end{document}